\documentclass[tighten,iop]{emulateapj}

\usepackage[english]{babel}
\usepackage{amsmath}
\usepackage{amsfonts}
\usepackage{amssymb}
\usepackage{graphicx}
\usepackage[dvips]{color} 

\usepackage{natbib}

\shorttitle{3D Boltzmann-Hydrodynamical Code: I SR Treatments}
\shortauthors{Nagakura et al.}

\begin{document}


\title{Three-Dimensional Boltzmann-Hydro Code for core-collapse in massive stars

 I. Special Relativistic Treatments}

\author{Hiroki Nagakura$^{1}$, Kohsuke Sumiyoshi$^{2}$ and Shoichi Yamada$^{3,4}$
}
\address{$^1$Yukawa Institute for Theoretical Physics, Kyoto
  University, Oiwake-cho, Kitashirakawa, Sakyo-ku, Kyoto, 606-8502,
  Japan}

\address{$^2$Numazu College of Technology, Ooka 3600, Numazu, Shizuoka 410-8501, Japan}

\address{$^3$Advanced Research Institute for Science \&
Engineering, Waseda University, 3-4-1 Okubo,
Shinjuku, Tokyo 169-8555, Japan}
\address{$^4$Department of Science and Engineering, Waseda
  University, 3-4-1 Okubo, Shinjuku, Tokyo 169-8555, Japan}

\begin{abstract}
We propose a novel numerical method for solving multi-dimensional, special relativistic Boltzmann equations for neutrinos coupled to hydrodynamics equations. It is meant to be applied to simulations of core-collapse supernovae. We handle special relativity in a non-conventional way, taking account of all orders of $v/c$. Consistent treatment of advection and collision terms in the Boltzmann equations is the source of difficulties, which we overcome by employing two different energy grids: {\it Lagrangian remapped} and {\it laboratory fixed} grids. We conduct a series of basic tests and perform a one-dimensional simulation of core-collapse, bounce and shock-stall for a $15 M_{\sun}$ progenitor model with a minimum but essential set of microphysics. We demonstrate in the latter simulation that our new code is capable of handling all phases in core-collapse supernova. For comparison, a non-relativistic simulation is also conducted with the same code, and we show that they produce qualitatively wrong results in neutrino transfer. Finally, we discuss a possible incorporation of general relativistic effects in our method.
\end{abstract}


\section{Introduction}\label{sec:intro}
 Quantitative studies on the mechanism of core-collapse supernovae (CCSNe) require detailed numerical simulations. Except for low-mass ($8 \sim 10 M_{\sun}$) progenitors, elaborate one-dimensional (1D) simulations under spherical symmetry have not reproduced the supernova explosion \citep{2005ApJ...629..922S,2005ApJ...620..840L,2006A&A...450..345K,2007AIPC..937..370B}. Last decade, most of supernova modelers have focused on the multi-dimensional (Multi-D) aspects of dynamics
 (see e.g., \citet{2012AdAst2012E..39K,2012ARNPS..62..407J,2013RvMP...85..245B} for recent review). In the post-bounce phase, instabilitities drive post-shock accretion flows into turbulence, making dynamics intrinsically multi-D. This may be crucial for the supernova explosion, since the non-spherical turbulent motions increase the dwell time of material in the gain region, enhancing its absorption of hot neutrinos, boosting the post shock pressure, and eventually pushing the shock wave outwards \citep{2012ApJ...749...98T,2013ApJ...765..110D}.

 As a matter of fact, we have recently witnessed shock revival in some of the currently most advanced simulations \citep{2006ApJ...640..878B,2009ApJ...694..664M,2010PASJ...62L..49S,2012nuco.confE.208L,2012ApJ...761...72M,2012ApJ...756...84M,2013arXiv1308.5755T}, which has raised our hope that we will finally unveil the mechanism of CCSNe. Unfortunately, however, success or failure of the supernova explosion is a delicate problem. In fact, the latest results of Multi-D simulations by different groups are still at odds with one another and no consensus has yet emerged concerning which ingredient(s) is (are) essential for explosion. Although various approaches, both phenomenological and ab initio, are being undertaken at present, only better simulations possibly with a Boltzmann-equation solver that incorporate detailed microphysics and general relativity (GR) may give the conclusive answer.

Towards this goal, we are developing a numerical code for neutrino transfer, which solves the Boltzmann equations \citep{2012ApJS..199...17S}. Our code is based on the discrete-ordinate $S_{n}$ method, which finite-differences the Boltzmann equations, deploying multi-angle and multi-energy bins in momentum space. Using some snapshots from three-dimensional (3D) supernova simulations, \citet{2012ApJS..199...17S} demonstrated the capabilities of this new code, which implements the minimum set of neutrino reactions (see also \citet{2014arXiv1403.4476S}). These simulations concerned neutrino transfer in static backgrounds, however, and no back-reactions to matter were taken into account.

 The next step should be a coupling of this code with a hydrodynamical code. This may not be so simple, though. Spherically symmetric 1D computations may be easier, since they can adopt Lagrangian formulations both for neutrino transfer and hydrodynamics \citep{1993ApJ...405..669M,2001PhRvL..86.1935M,2005ApJ...620..840L,2005ApJ...629..922S,2007ApJ...667..382S}. Such formalisms as they are could not be applied in Multi-D, however, and different formulations should be developed for the Multi-D Boltzmann-Hydro simulations, i.e. the simulations that solve the Boltzmann equations and hydrodynamical equations simultaneously in multi-dimensions.

Unlike the previous 1D codes, we adopt an Eulerian picture in this paper. There are several reasons for this choice. Among other things, we have in mind that the Boltzmann solver will be coupled with a Multi-D Eulerian hydrodynamics and gravity solvers, which have been well established and widely used in the high-energy astrophysical community. In addition, the Eulerian picture has a benefit to easily handle the left hand side of Boltzmann equation, i.e., advection terms. In general, Lagrangian formulations need to treat derivatives with respect to neutrino energy, which correspond to the Doppler's effect caused by spatial and/or temporal variations in fluid velocities. This may cause problems particularly at a shock wave, where fluid velocities are discontinuous. For these reasons, we have opted the Eulerian approach.


It should be noted, however, that the Eulerian approach has its own demerits. For example, we normally need to handle transformations between the laboratory frame and the fluid-rest frame defined locally, which are nothing but Lorentz transformations for flat space-time, since neutrino-matter interactions are best described in the fluid-rest frame. Physically consistent treatments of both advection and collision terms in the Multi-D Boltzmann solver are technically difficult particularly in the $S_{n}$ method, since special relativistic (SR) effects such as Doppler shifts and aberrations should be handled on a rather coarse grid in the momentum space (see Section~\ref{sec:difficulty} for more details). 
 Previous studies have attempted to alleviate this difficulty by employing an expansion of basic equations up to $\mathcal{O}(v/c)$ (see e.g., \citet{2007ApJ...659.1458H}). In CCSNe, the maximum fluid velocity is around $10 \%$ of the speed of light and such a first-order approximation may be justified. The resultant equations are fairly complex and not easy to treat numerically, however, and the formulation is certainly not applicable to highly relativistic phenomena.

It should be also mentioned that several groups \citep{2005PhRvD..72d3007C,2013PhRvD..88b3011C,2013arXiv1307.1666P} are developing different formulations for Multi-D Boltzmann-Hydro simulations, which are yet to be implemented. \citet{2008ApJ...685.1069O} performed detailed 2D Boltzmann-Hydro supernova simulations in the post bounce phase but they ignored SR effects. As shown later, consistent treatments of SR effects are indispensable to obtain correct behaviors in neutrino transfer. They are also the first step towards fully GR Boltzmann simulations, which will be needed to study more extreme phenomena such as black hole formations.

In this paper, we propose a novel formulation for the numerical computations of Multi-D SR Boltzmann transfer based on the $S_{n}$ method, which treats SR effects to all orders of $v/c$, where $c$ and $v$ denote the speed of light and fluid velocity, respectively. The accuracy of our new method is checked by the standard tests as well as by a realistic simulation of spherical collapse of a $15 M_{\sun}$ progenitor. As explained in the next section, the appropriate treatment of SR effects is crucial for numerically capturing the neutrino-trapping and the subsequent evolution up to bounce. In this paper, we particularly focus on this issue, and more detailed quantitative analyses of realistic supernova simulations by our Boltzmann-Hydro code will be reported later separately.



This paper is organized as follows. To facilitate readers' understanding, we first give intuitive arguments on the importance of SR effect from the perspective of phase space (in Section~\ref{sec:neutrinotrap}), which will make clear why non-relativistic Boltzmann-Hydro simulations fail to capture neutrino advections with matter and yield qualitatively wrong distributions of neutrinos. In Section~\ref{sec:difficulty}, it is also emphasized that the treatment of SR effects is not so easy practically, and it is explained what is the main obstacle. Then the basic equations and formulations are presented in Section~\ref{sec:SRBeqneutri}. After introducing two independent energy-grids (which are essential for our SR treatment) in Section~\ref{sec:twoenegrids}, the numerical algorithms are given in Section~\ref{sec:numeimple}. We examine the accuracy of our new method by a series of SR Boltzmann and Boltzmann-Hydro simulations in Section~\ref{sec:validation}. Finally we conclude the paper with a summary and discussions on further extensions of our code to a GR version in Section~\ref{sec:summary}. Throughout this paper, Greek and Latin subscripts denote space-time and space components, respectively. We use the metric signature of $- + + +$. Unless otherwise stated, we work in units with $c=G=1$, where $G$ is the gravitational constant.

\section{SR effects and neutrino trapping} \label{sec:neutrinotrap}

\begin{figure*}
\vspace{15mm}
\epsscale{0.5}
\plotone{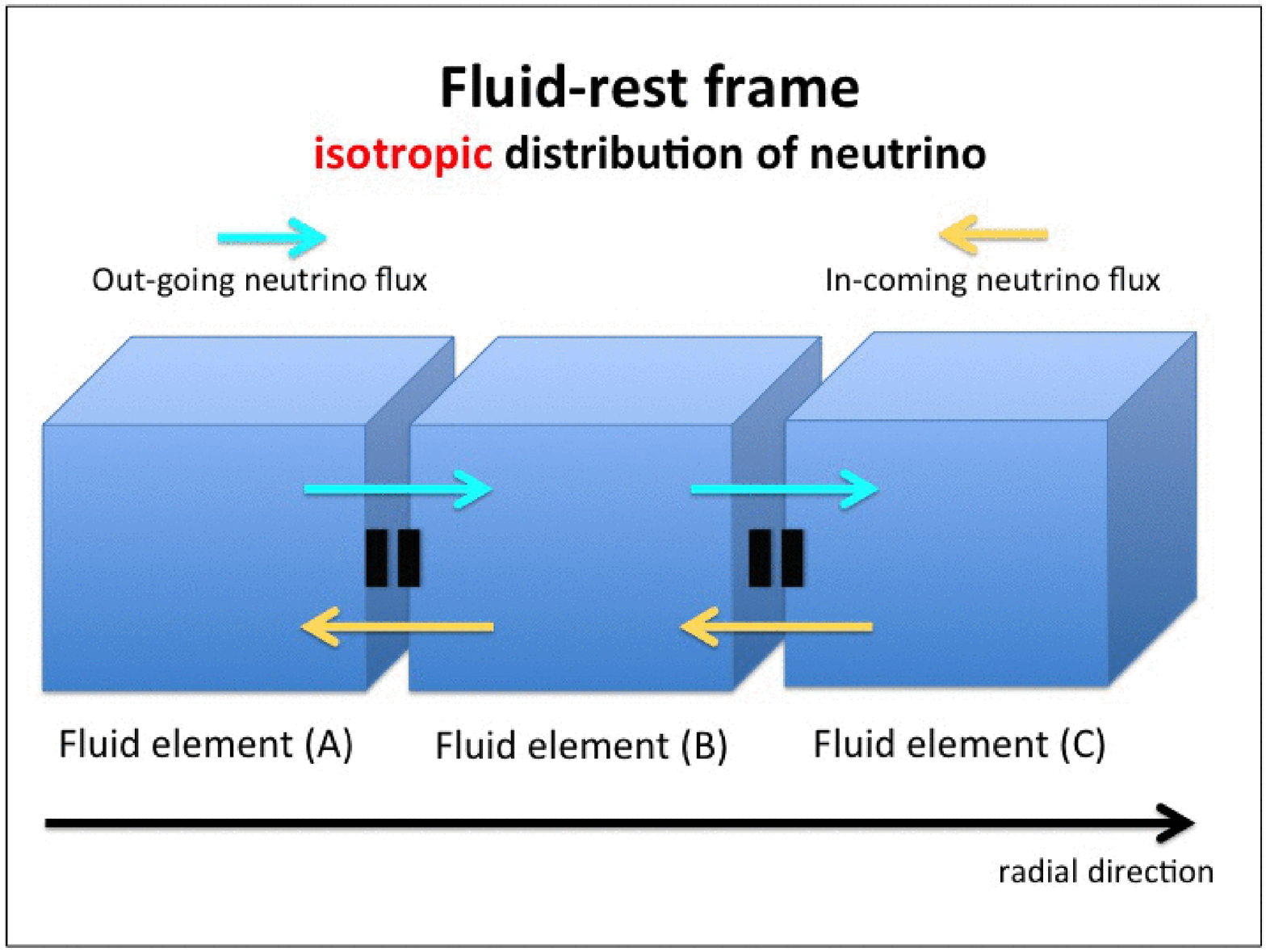}
\plotone{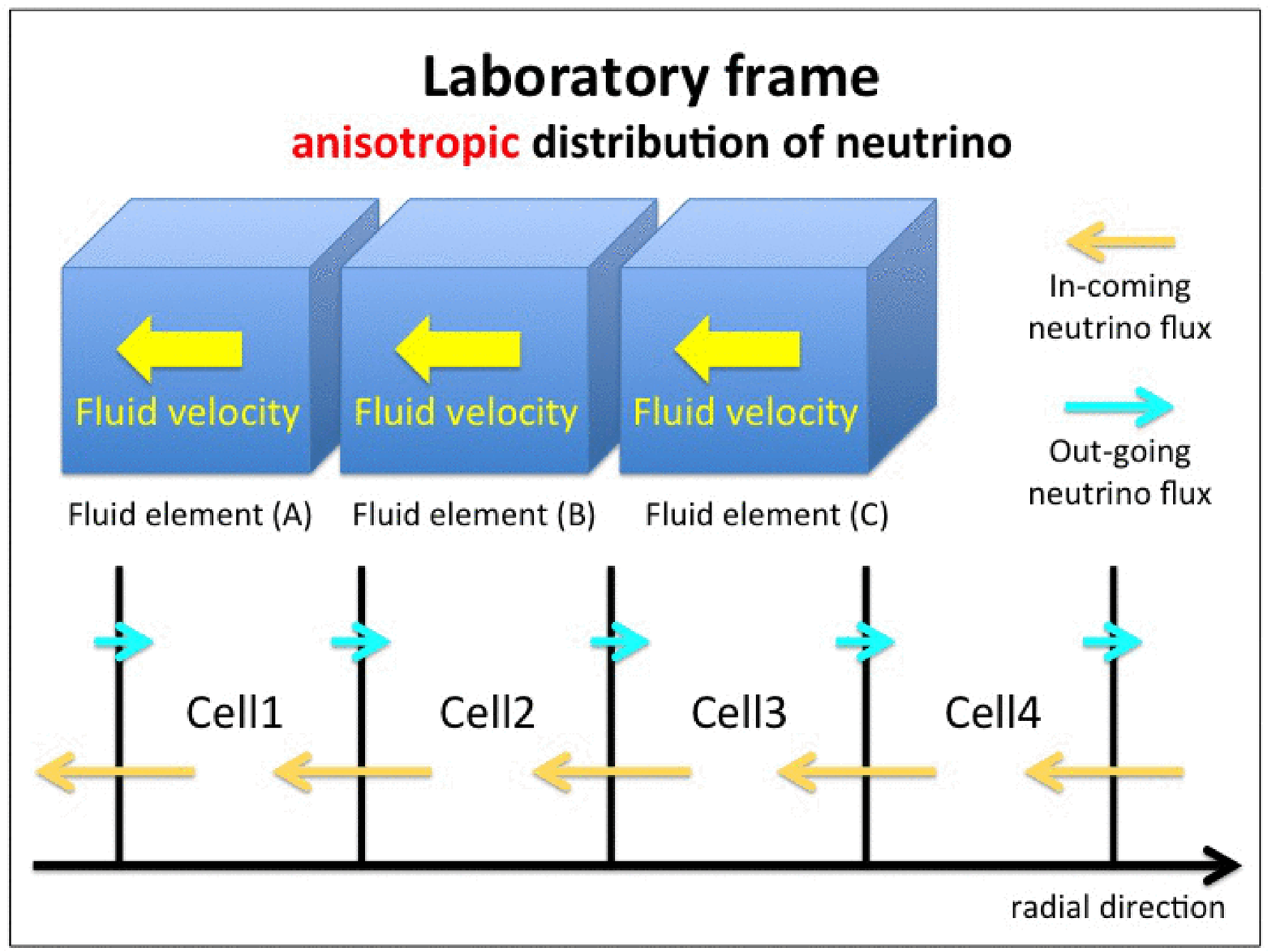}
\caption{Left panel: In-coming and out-going neutrino fluxes at the interface between fluid elements, which are measured on the fluid-rest frame. Right panel: The neutrino fluxes are displayed in the laboratory frame, in which matter is moving inwards. The bottom picture shows the in-coming (yellow) fluxes are larger than the out-going (blue) ones at the interfaces of laboratory-fixed spatial grids.
\label{schema1}} 
\end{figure*}

Before going to details of our SR Boltzmann formulation and its numerical algorithm, we first give intuitive arguments on the importance of SR effects. As will be observed below, the ignorance of SR effects yields qualitatively wrong behaviors in neutrino distributions. The key ingredient is the angular distribution in phase space: isotropic distributions in the fluid-rest frame become {\it anisotropic} in the laboratory frame after Lorentz transformations, the fact that ensures advection of neutrinos with matter and eventually the neutrino trapping.


In the following, we will explain this in a simplified and idealized set-up. We consider neutrino transfer in moving matter that has uniform velocity and thermodynamic quantities. We assume in addition that neutrinos and matter are strongly coupled with each other via scattering and, as a result, the neutrino distribution function in phase space, $f$, is isotropic in the fluid-rest frame. This is a situation similar to the ones we see locally in the neutrino trapping phase. Then the neutrino flux at each point vanishes in the fluid-rest frame and there is no net flux traversing fluid elements (see the left panel in Figure~\ref{schema1}). The neutrino number in each fluid element is conserved as the fluid element moves at the finite velocity. This is the advection of neutrinos with matter and it should be evident why the Lagrangian approach is advantageous in dealing with it.

For comparison, the right panel in Figure~\ref{schema1} describes the same situation except in the laboratory frame. Here, we assume that the fluid is advected inwards (or leftwards in the figure). Since the neutrinos should be advected in the same direction as the fluid in the laboratory frame, the in-coming neutrino flux is larger than the out-going one, which means that the angular distribution of neutrinos is {\it anisotropic} in this frame. From the SR point of view, such anisotropies arise from the Doppler shift and relativistic beaming by Lorentz transformations. The mathematical expression of SR Boltzmann equations will be given in Section~\ref{sec:SRBeqneutri}.

If we neglected all SR effects, not distinguishing the laboratory and fluid-rest frames, we would not obtain the neutrino advection with matter, which is crucial for the neutrino trapping in the collapsing phase. In fact, neutrinos would be left behind as fluids are advected. The supernova core is not homogeneous in reality and both matter and neutrino densities are highest at the center. In the absence of advection, neutrinos would always flow outwards when they actually should move inwards, keeping pace with matter, and be effectively trapped in the core. As we will show later in Section~\ref{subsec:1DSRBoltz_Hydro}, the number density of electron-type neutrinos becomes significantly smaller near the center for non-relativistic simulations. It affects in turn the evolution of electron fraction and the size of inner core and eventually all the supernova dynamics thereafter.



\section{Difficulties in handling SR effects} \label{sec:difficulty}

\begin{figure*}
\vspace{15mm}
\epsscale{1.0}
\plotone{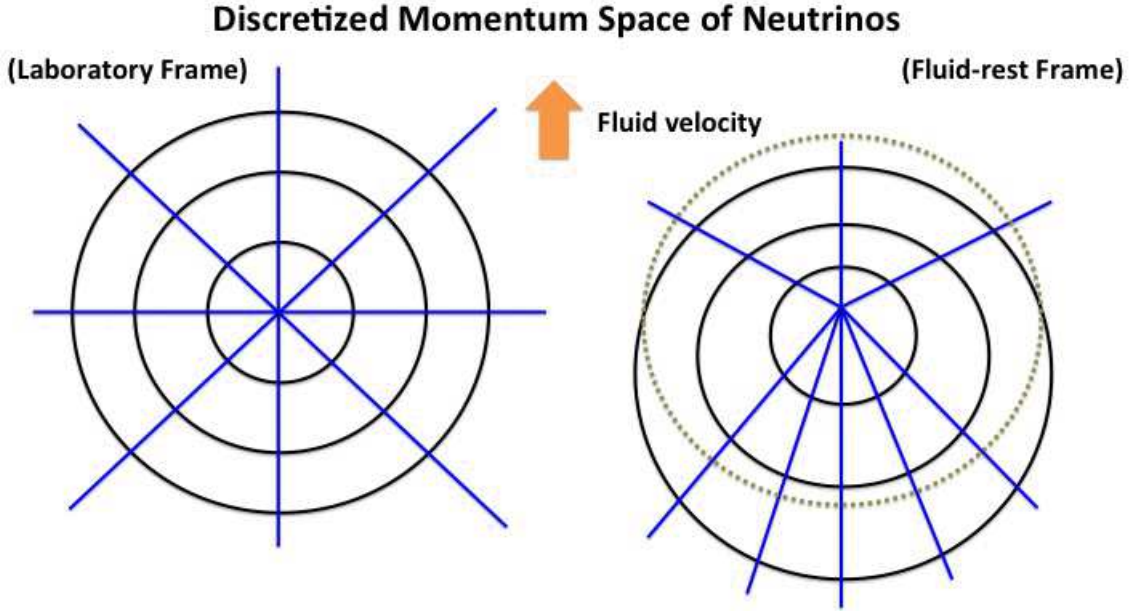}
\caption{Left: Discretized momentum space of neutrinos in the laboratory frame. Spherical coordinates are employed. The radial direction corresponds to neutrino energy and the azimuthal dimension is omitted. The grid in each dimension may not be uniform. Right: The Lorentz-transformed mesh in the fluid-rest frame. The blue lines correspond to the radial lines whereas the black lines are transformed from the concentric circles in the left panel. The brown dots show an isoenergy circle in the fluid-rest frame for comparison. Matter is assumed to move upward in this figure.
\label{momspacedist}} 
\end{figure*}

We give intuitive explanations more in detail on why SR treatments are not easy in the $S_{n}$ method, which we employ in this paper. The main source of the difficulties is scatterings, particularly those between neutrinos and nucleons (and nuclei). Other reactions such as neutrino absorptions and emissions have no technical difficulties\footnote{Of course, non-isoenergetic scatterings on electrons and neutrinos and pair processes are another complication, which will be addressed in the future.}. We hence focus only on the isoenergetic scatterings in this section.

As mentioned in the previous sections, our Boltzmann-Hydro code is based on the Eulerian picture, and we discretize 6D phase space in the laboratory frame, as shown in the the left panel in Figure~\ref{momspacedist}. In this picture, spherical coordinates in momentum space are adopted with the azimuthal dimension being collapsed. The radial direction corresponds to neutrino energy. Although the picture is drawn that way, gridding in each dimension is not necessarily uniform.

We first consider the isoenergetic scattering under the condition that fluid is at rest and, as a consequence, the laboratory fame coincides with the fluid-rest frame. When a neutrino undergoes the isoenergetic scattering, it changes its flight direction specified by two angles, preserving energy. In the discretized momentum space, the neutrino moves from a bin to another with the same radial-grid number. The important thing is that only the angular-grid number is changed. In this case, there is no difficulty and, indeed, it has been implemented in \citet{2012ApJS..199...17S,2014arXiv1403.4476S}.

In the presence of non-vanishing fluid velocities, the problem becomes qualitatively different. In this case, the laboratory frame is different from the fluid-rest frame and they are related with each other via a Lorentz transformation. The point is that the Lorentz transformation induces changes in both energy and angles. These energy shift and aberration are determined by the Doppler factor, which depends on the fluid velocity and neutrino angles (see Section~\ref{sec:SRBeqneutri}). This is most clearly demonstrated in the right panel of Figure~\ref{momspacedist}, in which the spherical coordinates given in the laboratory frame are Lorentz-transformed to the fluid-rest frame. It is evident that they are no longer spherically symmetric and distorted in the latter frame. This picture summarizes the difficulties in the treatment of scatterings even if they are isoenergetic. As is well known, the neutrino distribution function $f$ is a Lorentz invariant and its values at corresponding points in different frames are identical. The important point, however, is the fact that grid points are shifted by Lorentz transformations and concentric (equivalently isoenergetic) spheres in the laboratory frame are no longer spheres in the fluid-rest frame. As a consequence, some interpolations are inevitable in evaluating the collision terms for scatterings in the fluid-rest frame if one were to avoid the $v/c$ expansion. There are several difficulties to carry out this interpolation particularly in neutrino energy, though. The reasons are described shortly.

The rather low energy resolution we can afford in the Boltzmann code is one of the reasons. We can deploy at most $\sim 20$ energy bins (see \citet{2012arXiv1205.6284K}). The distribution function $f$ depends strongly on the neutrino energy in general. In particular, it decreases almost exponentially at high energies. On the numerical mesh, $f$ may change several orders of magnitude between adjacent energy-grid points. Highly accurate interpolations of $f$ are hence required on the coarse mesh. Note that since the isoenergetic scatterings between neutrinos and nucleons and/or nuclei dominate other reactions in CCSNe, the time step ($\Delta t$) of simulations is mostly determined by these processes. If the interpolations of $f$ are not accurate at high energies, we might find that $\Delta t$ becomes unreasonably small because of a large number of artificial scatterings. The fact that high energy neutrinos have larger cross sections makes matter worse. Not to mention, in the interpolation we further have to care about the conservation of neutrino numbers in scatterings.

After giving the SR Boltzmann equations in the next section, we present our idea to overcome these difficulties. We then demonstrate our successful handling of the isoenergetic scatterings in the realistic supernova simulations (see Section~\ref{sec:validation}).

\section{SR Boltzmann equations for neutrinos} \label{sec:SRBeqneutri}

\begin{figure*}
\vspace{15mm}
\epsscale{0.5}
\plotone{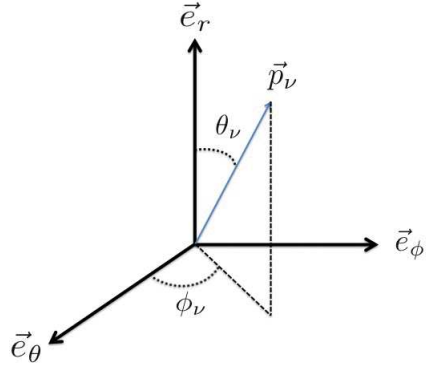}
\caption{Local orthonormal bases that measure neutrino momentum. ${\vec{e}_{r}}$, ${\vec{e}_{\theta}}$, and ${\vec{e}_{\phi}}$ are aligned with the spatial spherical coordinates as the subscripts show.
\label{neutrinomompic}} 
\end{figure*}

 We start with the covariant form of Boltzmann equation:
\begin{eqnarray}
p^{\mu} \frac{\partial f}{\partial x^{\mu}}
+ \frac{dp^{i}}{d\tau} \frac{\partial f}{\partial p^{i}}
= \Bigl( \frac{\delta f}{\delta \tau} \Bigr)_{\rm col},  \label{eq:basicBoltz}
\end{eqnarray}
which is valid even in curved space-time.
In the above expression, $f$($= f(x^{\mu},p^{i})$) denotes the neutrino distribution function in phase space; $x^{\mu}$ and $p^{\mu}$ are space-time coordinates and four-momentum of neutrino, respectively; since the latter satisfies the on-shell condition: $p^{\mu} p_{\mu} = - m_{\nu}^2$, in which $m_{\nu}$ is a neutrino mass, only three of four components are independent and this is why only spatial components appear in the second term on the left hand side; $\tau$ stands for the affine parameter of neutrino trajectory. The left hand side of Eq.~(\ref{eq:basicBoltz}) expresses a geodesic motion in the phase space, while the right hand side denotes symbolically the so-called collision terms, i.e., the terms that give the rate of changes in $f$ due to neutrino-matter interactions.

On the spherical coordinates in flat space-time, which are the coordinates we employ for the laboratory frame in our Eulerian approach, Eq.~(\ref{eq:basicBoltz}) is cast into the following conservation form:
\begin{eqnarray}
&& \frac{\partial f}{\partial t} 
+ \frac{\mu_{\nu}}{r^{2}} \frac{\partial}{\partial r} (r^{2} f)
+ \frac{\sqrt{1-\mu_{\nu}^{2}}~{\rm cos}~\phi_{\nu}}{r {\rm sin}~\theta} 
  \frac{\partial}{\partial \theta} ({\rm sin}~\theta f) \nonumber \\
&& + \frac{\sqrt{1-\mu_{\nu}^{2}}~{\rm sin}~\phi_{\nu}}{r {\rm sin}~\theta} 
  \frac{\partial f}{\partial \phi}
+ \frac{1}{r} 
  \frac{\partial}{\partial \mu_{\nu}} [(1-\mu_{\nu}^{2}) f] \nonumber \\
&& - \frac{\sqrt{1-\mu_{\nu}^{2}}}{r} 
  \frac{{\rm cos}~\theta}{{\rm sin}~\theta}
  \frac{\partial}{\partial \phi_{\nu}} ({\rm sin}~\phi_{\nu} f)
= \Bigl( \frac{\delta f}{\delta t} \Bigr)_{{\rm col}}^{{\rm lb}} ,
 \label{eqn:eqtransfin-spherical}
\end{eqnarray}
where $r$, $\theta$, $\phi$ denote the spatial variables; as three independent components of neutrino four-momentum, we do not use its spacial components but adopt energy and two angles, $\theta_{\nu}$ and $\phi_{\nu}$ (see Figure~\ref{neutrinomompic}); $\mu_{\nu}$ is defined as $\mu_{\nu} \equiv \cos{\theta_{\nu}}$. In Eq.~(\ref{eqn:eqtransfin-spherical}) and the rest of this paper, we assume that neutrinos are massless, which is well justified as long as neutrino oscillations are ignored.


The collision term in Eq.~(\ref{eqn:eqtransfin-spherical}), which is expressed with the laboratory time $t$, is related with the original collision term in equation~(\ref{eq:basicBoltz}) as
\begin{eqnarray}
\Bigl( \frac{\delta f}{\delta \tau} \Bigr)_{\rm col}
= \varepsilon^{\rm{lb}} \Bigl( \frac{\delta f}{\delta t} \Bigr)_{{\rm col}}^{\rm{lb}}, \label{eq:colintocova}
\end{eqnarray}
where $\varepsilon^{{\rm lb}}(\equiv p^{t})$ denotes the neutrino energy measured in the laboratory frame. Similarly, the collision term in the fluid-rest frame can be expressed with the proper time of each fluid element ($\tilde{t}$) as
\begin{eqnarray}
\Bigl( \frac{\delta f}{\delta \tau} \Bigr)_{\rm col}
= \varepsilon^{\rm{fr}} \Bigl( \frac{\delta f}{\delta \tilde{t}} \Bigr)_{{\rm col}}^{\rm{fr}}, \label{eq:colcotocova}
\end{eqnarray}
where $\varepsilon^{\rm{fr}}(\equiv p^{\tilde{t}} \equiv - u_{\mu} p^{\mu})$ denotes the neutrino energy in the fluid-rest frame. Here $u^{\mu}$ is the four-velocity of matter.

The Lorentz transformation of four-momentum gives the relation of neutrino energies in the fluid-rest and laboratory frames as
\begin{equation}
\varepsilon^{\rm{fr}} = \varepsilon^{{\rm lb}} \gamma 
(1 - \mbox{\boldmath $n$}^{{\rm lb}} \cdot \mbox{\boldmath $v$}) , \label{eqn:Lorentz-energy}
\end{equation}
where {\boldmath $v$}, $\gamma (\equiv (1-v^2)^{-1/2})$ denote the three-velocity and corresponding Lorentz factor of matter and $\mbox{\boldmath $n$}^{{\rm lb}}$ is the unit vector that indicates the flight direction of neutrino in the laboratory frame. The factor $D^{{\rm lb}} \equiv \gamma (1 - \mbox{\boldmath $n$}^{{\rm lb}} \cdot \mbox{\boldmath $v$})$ in Eq.~(\ref{eqn:Lorentz-energy}) expresses the Doppler shift of neutrino energy. From Eqs.~(\ref{eq:colintocova}) - (\ref{eqn:Lorentz-energy}), we can obtain the relation between the collision terms in the two frames as
\begin{eqnarray}
\Bigl( \frac{\delta f}{\delta t} \Bigr)_{{\rm col}}^{{\rm lb}} = 
D^{{\rm lb}} \Bigl( \frac{\delta f}{\delta \tilde{t}} \Bigr)_{{\rm col}}^{\rm{fr}}.
 \label{eq:collisionrela}
\end{eqnarray}

The Lorentz transformation also gives the relation between the flight directions in the fluid-rest and laboratory frames as
\begin{eqnarray}
\varepsilon^{\rm{fr}} \mbox{\boldmath $n$}^{\rm{fr}} = 
\varepsilon^{{\rm lb}}
[\mbox{\boldmath $n$}^{{\rm lb}} + \{ - \gamma + (\gamma - 1) 
\frac{\mbox{\boldmath $n$}^{{\rm lb}} \cdot \mbox{\boldmath $v$}}{v^2}\}
\mbox{\boldmath $v$}]. \label{eq:energytrans} 
\end{eqnarray}
Here $\mbox{\boldmath $n$}^{\rm{fr}}$ denotes the unit vector that specifies the flight direction of neutrino in the fluid-rest frame. Using the Doppler factor $D^{{\rm lb}}$, we obtain
\begin{eqnarray}
\mbox{\boldmath $n$}^{\rm{fr}} = 
\frac{1}{D^{{\rm lb}}}
[\mbox{\boldmath $n$}^{{\rm lb}} + \{ - \gamma + (\gamma - 1) 
\frac{\mbox{\boldmath $n$}^{{\rm lb}} \cdot \mbox{\boldmath $v$}}{v^2}\}
\mbox{\boldmath $v$}]. \label{eqn:Lorentz-direction}
\end{eqnarray}
Note that this relation no longer contains neutrino energy and the angle-transformations are decoupled from the energy transformations. This is a great simplification, which we make full use of in the following, and is a consequence of the assumption that neutrinos are massless. The solid-angle element is then transformed as
\begin{eqnarray}
d \Omega^{\rm{fr}} = \frac{1}{(D^{{\rm lb}})^2} d \Omega^{{\rm lb}}. \label{eqn:solidrela}
\end{eqnarray}


In the Boltzmann equation, neutrino-matter interactions are described in the collision terms. As is well known, they are obtained most easily in the fluid-rest frame. We hence evaluate the collision term in this frame and use Eq.~(\ref{eq:collisionrela}) to obtain the expression in the laboratory frame. The interactions that we take into account in this paper are the same as those in \citet{2012ApJS..199...17S}, the minimum set for supernova simulations. Since \citet{2012ApJS..199...17S} worked in the Newtonian approximation, we need the following replacements to employ their collision terms:
\begin{eqnarray}
 \Bigl[ \frac{1}{c} \frac{\delta f}{\delta t} \Bigr] &\rightarrow&
\Bigl( \frac{\delta f}{\delta \tilde{t}} \Bigr)_{{\rm col}}^{\rm{fr}}, \nonumber \\
 \varepsilon &\rightarrow& \varepsilon^{\rm{fr}}, \nonumber \\
 \Omega &\rightarrow& \Omega^{\rm{fr}}, \nonumber \\
 R_{*} &\rightarrow& R_{*}^{\rm{fr}}, \label{eq:replacement}
\end{eqnarray}
where $R_{*}$ denotes reaction kernels.

Here we take the collision terms for the isoenergetic scatterings in the laboratory frame and see how the neutrino-number conservation is ensured, which will be useful in the next section. Following \citet{2012ApJS..199...17S} and implementing the above replacements, we can write them as
\begin{eqnarray}
\Bigl( \frac{\delta f}{\delta \tilde{t}} \Bigr)_{scat}^{\rm{fr}} && (\varepsilon^{\rm{fr}},\Omega^{\rm{fr}}) = - \frac{(\varepsilon^{\rm{fr}})^{2}}{(2 \pi)^3} \int d \Omega^{{\rm 'fr}} R_{scat}^{\rm{fr}} (\Omega^{\rm{fr}},\Omega^{{\rm 'fr}}) \nonumber \\
&& \times \Bigl( f^{\rm{fr}} (\varepsilon^{\rm{fr}},\Omega^{\rm{fr}}) - f^{\rm{fr}} (\varepsilon^{\rm{fr}},\Omega^{{\rm 'fr}}) \Bigr), \label{eq:isoenergyscatkenel}
\end{eqnarray}
where $R_{scat}^{\rm{fr}}$ and $f^{\rm{fr}}$ denote the isoenergetic scattering-kernel and neutrino distribution function $f$ in the fluid-rest frame, respectively. The integration of Eq.~(\ref{eq:isoenergyscatkenel}) over the solid angle $\Omega^{\rm{fr}}$ vanishes due to symmetric properties of scattering kernel: $R(\Omega, \Omega^{'}) = R(\Omega^{'}, \Omega)$. This represents the conservation of neutrino number for the isoenergetic scatterings at each energy in the fluid-rest frame.

\section{Two energy-grids}\label{sec:twoenegrids}

\begin{figure*}
\vspace{15mm}
\epsscale{1.0}
\plotone{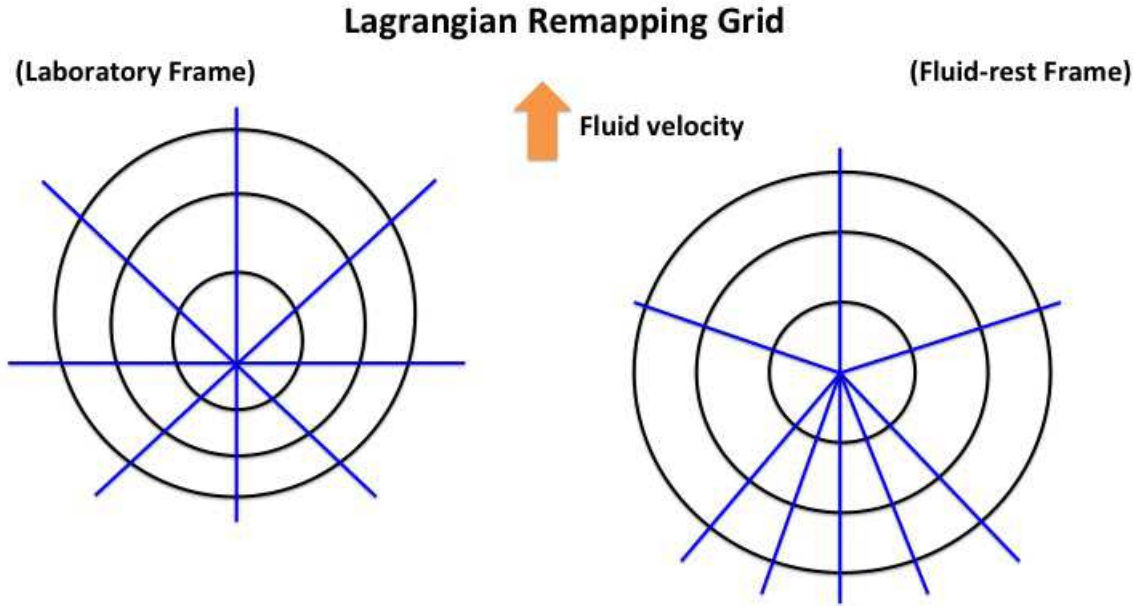}
\caption{Lagrangian remapped grid in the laboratory frame (left panel) and the Lorentz-transformed grid in the fluid-rest frame (right panel). The energy grid is isotropic in the fluid-rest frame whereas it becomes anisotropic in the laboratory frame. The angular grid, on the other hand, is uniform in the laboratory frame.
\label{momspacedist_laremap}} 
\end{figure*}

\begin{figure*}
\vspace{15mm}
\epsscale{1.0}
\plotone{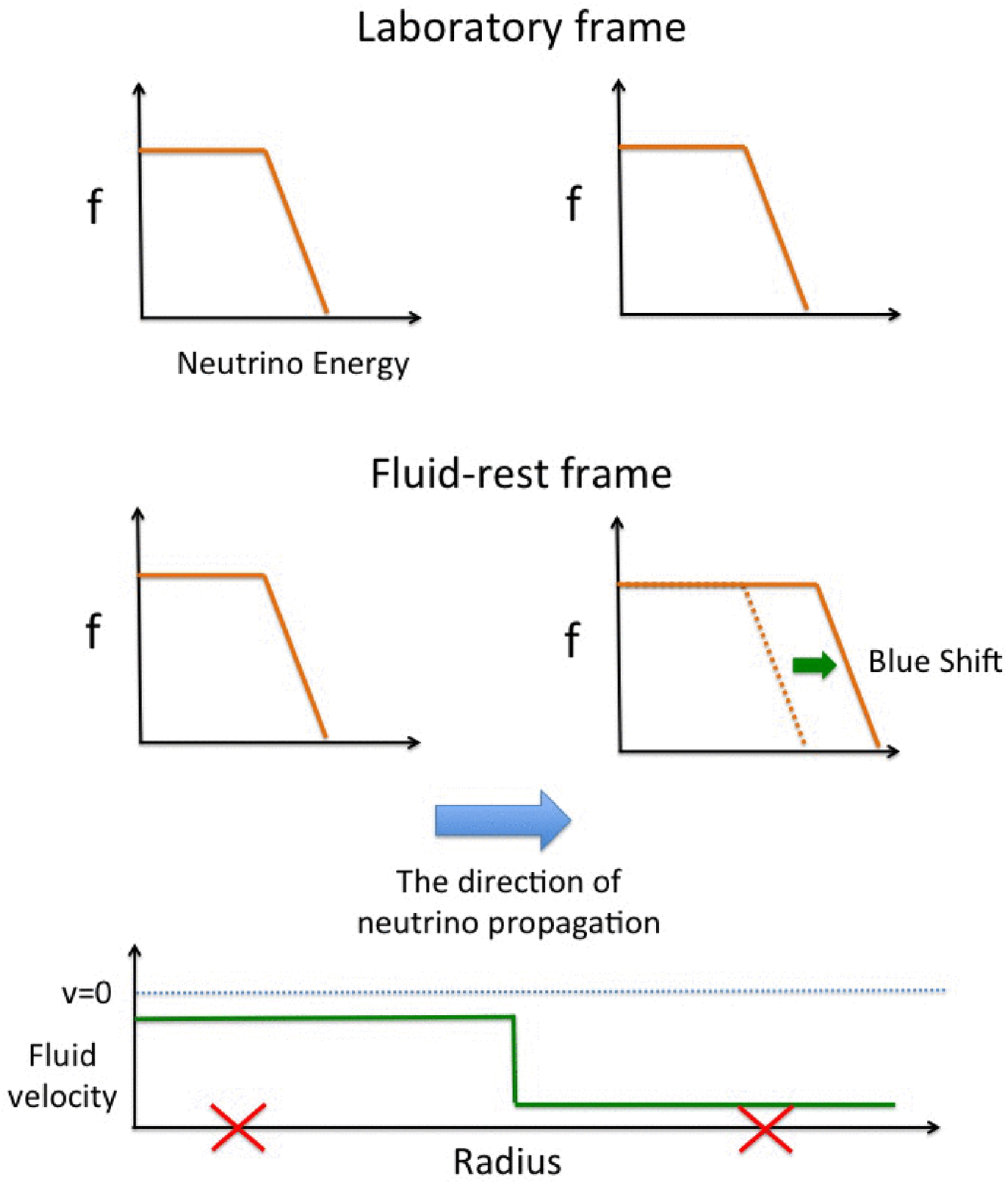}
\caption{Schematic pictures of the energy spectra of out-going neutrinos in the laboratory (upper) and fluid-rest frames (middle). Matter is assumed to be optically thin and flows inwards at piecewise constant velocities with a discontinuity in the middle (lower picture). The two red crosses in the bottom picture are locations where we measure the neutrino spectra. The spectrum should be unchanged across the discontinuity in the laboratory frame whereas it will be blue-shifted in the fluid-rest frame.
\label{Lagremap_com_Labo}} 
\end{figure*}

The origin of difficulties in the SR treatments is the fact that the neutrino momentum space is distorted by Lorentz transformations, i.e., the isoenergy surfaces in the laboratory frame do not coincide with the counterparts in the fluid-rest frame. We then need highly accurate interpolations in energy of $f$, taking care of neutrino-number conservation, whose difficulties in the $S_{n}$ method were elucidated in Section~\ref{sec:difficulty}.


 We overcome these difficulties by employing not the grid shown in the left panel of Fig.~\ref{momspacedist} but the so-called Lagrangian remapped grid (hereafter {\it LRG}) in the laboratory frame, which is Lorentz-transformed from the fluid-rest frame. It is emphasized that {\it LRG} is the one we mainly use in our Eulerian approach. In Figure~\ref{momspacedist_laremap}, we display the schematic picture of our {\it LRG} (see also Figure~\ref{momspacedist} for comparison). In this method, the energy grid is constructed so that it should be isotropic in the fluid-rest frame.

As a consequence, it becomes anisotropic in the laboratory frame as observed in the left panel. The energy grids obtained in the laboratory frame that way are different from point to point at each time and change also in time because of inhomogeneous fluid motions. Thanks to the isotropic energy grids in the fluid-rest frame, no special care is needed in the treatment of the isoenergetic scatterings on this grid. Note that the angular mapping is independent of energy. The angular grid is constructed, on the other hand, so that it should be uniform in the laboratory frame. It implies that the angular mesh is non-uniform in the fluid-rest frame as shown in the right panel.
 In contrast to the energy grid, the non-uniform angular grid in the fluid-rest frame causes no practical problems (see Eq.~(\ref{eq:omegainteg}) for the correction by angular aberration).


One may say that the Lagrangian remapping method is nothing but the canonical Lagrangian approach, but there are several differences between the two. One of the important differences lies in the treatment of advection terms on the left hand side of the Boltzmann equation. As we have already mentioned in Section~\ref{sec:intro}, the advection terms are fairly complicated in the Lagrangian approach due to the spatial inhomogeneities and temporal changes of matter velocity. We demonstrate this in a simplified situation in Figure~\ref{Lagremap_com_Labo}. Here, we consider neutrinos propagating outwards (or rightwards in the figure) in optically thin matter. We further assume that the matter is moving inwards (or leftwards in the figure) at velocities that are piecewise constant with $|v_{\rm{Left}}| < |v_{\rm{Right}}|$ (see the bottom panel in this figure). The discontinuity may be regarded as a standing shock wave.

In this situation, the neutrino-energy spectrum in the laboratory frame is uniform in space, since neutrinos are not interacting with matter at all (see the upper picture)\footnote{It is assumed here that the boundary condition is fixed and a steady state has been established.}. This is not the case in the fluid-rest frame, however. It is in fact blue-shifted at the discontinuity of matter velocity (see the middle picture in Figure~\ref{Lagremap_com_Labo}). In the Lagrangian picture, such energy shifts are expressed as the advection in energy space and given by the partial derivative with respect to energy on the left hand side of the Boltzmann equation. In the present case, there should be an infinite energy flux at the discontinuity\footnote{In numerical simulations, such a discontinuity is somewhat smeared and the flux becomes always finite.}. In our Lagrangian remapping method, on the other hand, we work in the laboratory frame, in which the energy grid is anisotropic as shown in the left panel of Figure~\ref{momspacedist_laremap} and the blue-shift in the spectrum is just compensated for by the contraction of the energy grid in the outward direction and, as a consequence, the energy spectrum is unchanged across the discontinuity (see the upper picture in Figure~\ref{Lagremap_com_Labo}).

It is easily understood that the use of {\it LRG}, which is anisotropic and spatially non-uniform, complicates the calculation of spatial and neutrino-angular advection, a problem similar to the one in the ordinary Lagrangian method. This is mitigated in our method, however, by the introduction of yet another energy grid, which is isotropic and identical at all spatial grid points in the laboratory frame (referred hereafter to as the laboratory-fixed grid or {\it LFG}; see also Section~\ref{subsec:step4} for more details). {\it LFG} is employed only to calculate the neutrino advection. Note that as long as we work in the laboratory frame, energy-derivative terms do not appear explicitly on the left hand side of the Boltzmann equation and the advection on the {\it LFG} is particularly simple.
 It should be repeated that {\it LFG} is a grid only for temporary use to treat the neutrino advection. Accordingly $f$ on the {\it LFG}, which is obtained by interpolation in our method, is also a temporal variable. Instead, $f$ on {\it LRG} is the quantity to be solved and stored in our code.


\section{Numerical Implementations}\label{sec:numeimple}

\begin{figure*}
\vspace{15mm}
\epsscale{1.0}
\plotone{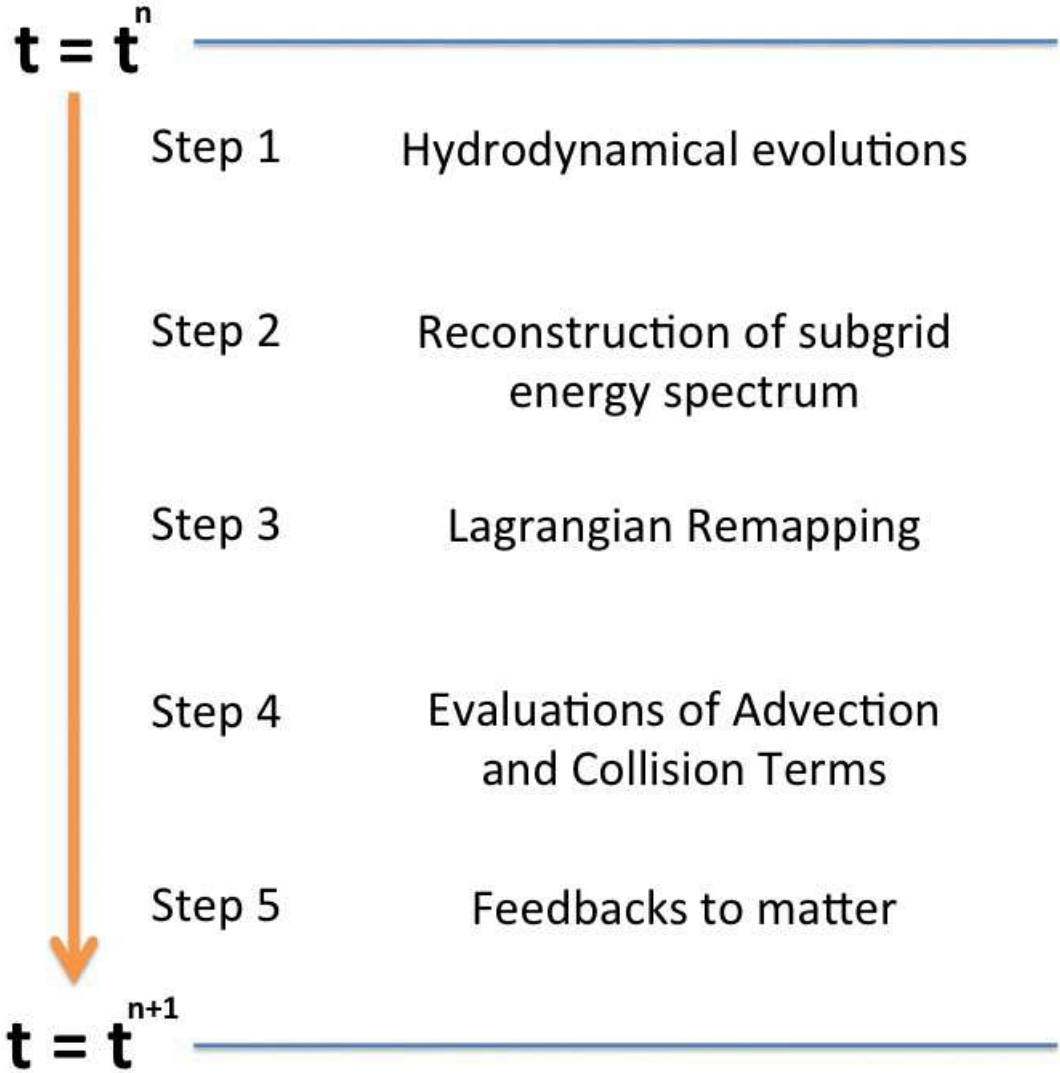}
\caption{The flow chart for our Boltzmann-Hydro solver.
\label{flowchartBoltzHydro}} 
\end{figure*}

In this section, we explain the detailed numerical algorithm to implement various elements described above to our Boltzmann-Hydro solver, paying particular attention to the usage of different energy grids. Figure~\ref{flowchartBoltzHydro} summarizes multiple steps needed to update a numerical solution from $t=t^{n}$ to $t^{n+1}$, where the superscripts represent the time steps. In the following, we describe each step in order in detail.


\subsection{Step~1: Hydrodynamical evolutions} \label{subsec:hdestep1}
In our Boltzmann-Hydro solver, the operator splitting is employed: we first compute hydrodynamics, neglecting neutrino interactions, i.e., in the adiabatic manner; then from Step~2 through Step~4 we perform neutrino transfer for the matter distribution given in the first step as described below; feedbacks from neutrino interactions to the internal energy, velocity and electron fraction of matter are taken into account in Step~5.

 The numerical code for the hydrodynamical evolution is essentially the same as that in \citet{2013ApJ...765..123N}. It is based on the so-called central scheme with an explicit time evolution \citep{2000JCoPh.160..241K,2008ApJ...689..391N,2011ApJ...731...80N}. The code was successfully applied to the simulations of Standing Accretion Shock Instability (SASI) in the post-bounce phase in our previous study \citep{2013ApJ...765..123N}. It is also noted that a series of standard tests for hydrodynamical schemes (e.g., shock tube problems) were carried out in \citet{2011ApJ...731...80N}.

Although our Boltzmann solver is fully SR, the hydrodynamics solver is Newtonian. As a matter of fact, it is fully general relativistic \citep{2008ApJ...689..391N} except for its gravity solver, which is Newtonian and based on the MICCG technique \citep{2011ApJ...731...80N}.  The implementation of an Einstein equation solver is currently underway, the perspective of which will be mentioned in Section~\ref{sec:summary}.


 The basic equations of Newtonian hydrodynamics in spherical coordinates are written in the following form:
\begin{eqnarray}
\partial_{t}{\mbox{\boldmath $Q$}} + \partial_{j}{\mbox{\boldmath $U^{j}$}} = \mbox{\boldmath $W_{h}$} + \mbox{\boldmath $W_{i}$}, \label{eq:hydroConservativeform}
\end{eqnarray}
where each term is given as
\begin{eqnarray}
&&\hspace{20mm} \mbox{\boldmath $Q$} =
\left(
\begin{array}{c}
\sqrt{g} \rho \\
\sqrt{g} \rho v_{r} \\
\sqrt{g} \rho v_{\theta} \\
\sqrt{g} \rho v_{\phi} \\
\sqrt{g} ( e + \frac{1}{2} \rho v^2) \\
\sqrt{g} \rho Y_{e}
\end{array}
\right), \\
&&\hspace{14mm} \mbox{\boldmath $U^{j}$} =
\left(
\begin{array}{c}
\sqrt{g} \rho v^{j} \\
\sqrt{g} (\rho v_{r} v^{j} + p \delta_r^{j})\\
\sqrt{g} (\rho v_{\theta} v^{j} + p \delta_{\theta}^{j})\\
\sqrt{g} (\rho v_{\phi} v^{j} + p \delta_{\phi}^{j})\\
\sqrt{g} ( e + p + \frac{1}{2} \rho v^2) v^{j} \\
\sqrt{g} \rho Y_{e} v^{j}
\end{array}
\right), \\
&&\mbox{\boldmath $W_{h}$} =
\left(
\begin{array}{c}
0 \\
\sqrt{g} \rho \Bigl( - \psi_{,r} + r (v^{\theta})^2 + r {\rm sin}^2\theta (v^{\phi})^2 + \dfrac{2p}{r \rho} \Bigr)\\
\sqrt{g} \rho \Bigl( - \psi_{,\theta} r^2 + {\rm sin}\theta {\rm cos}\theta (v^{\phi})^2 + \dfrac{p  {\rm cos}\theta }{ \rho {\rm sin}\theta }  \Bigr)\\
- \sqrt{g} \rho \psi_{,\phi} \\
- \sqrt{g} \rho v^{l} \psi_{,l} \\
0
\end{array}
\right), \\
&&\hspace{23mm} \mbox{\boldmath $W_i$} =
\left(
\begin{array}{c}
0 \\
- \sqrt{g} G^{r} \\
- \sqrt{g} G^{\theta} \\
- \sqrt{g} G^{\phi} \\
- \sqrt{g} G^{t} \\
- \sqrt{g} \Gamma
\end{array}
\right). \label{eq:systemeqcompo}
\end{eqnarray}
Note that $\mbox{\boldmath $W_{i}$}$ corresponds to the interactions between neutrinos and matter (the explicit expressions will be presented in Step~5) and $\sqrt{g} (= r^2 {\rm sin}\theta)$ denotes the volume factor in the spherical coordinates. Other variables, $\rho$, $p$, $e$, $Y_{e}$, $v^{j}$, and $\psi$, are the mass density, pressure, internal energy density, electron fraction, fluid velocity, and Newtonian gravitational potential, respectively.
 The Newtonian self-gravity is solved with the Poisson's equation,
\begin{eqnarray}
\Delta \psi = 4 \pi \rho. \label{eq:poissongra}
\end{eqnarray}

In our central scheme, the above system of equations is finite-differenced in space with the piecewise parabolic (PPM) interpolation and the total-variation diminishing (TVD) Runge-Kutta method is employed for time integration, which achieves second-order accuracy in both space and time.
We adopt the procedure proposed by \citet{2010ApJS..189..104M} in solving the energy equation (the 5th component in Eq.~(\ref{eq:hydroConservativeform})), which reduces secular errors in the energy conservation.

Throughout this paper, we use Shen's equation of state (EOS) \citep{2011ApJS..197...20S} with lepton and photon contributions being added. (see e.g., \citet{2013ApJ...765..123N}). The original EOS table is rather coarse for the simulation of CCSNe. Indeed, we have found that tri-linear interpolations in the original table reduce the accuracy of simulations particularly at the transition from inhomogeneous to homogeneous nuclear matter. We have hence reconstructed a new EOS table by interpolating all quantities with the tricubic Hermite functions. It is several times finer in $\rho, Y_{e}$ and $T$ than the original table.

\subsection{Step~2: Reconstruction of subgrid energy spectrum} \label{subsec:espestep2}

\begin{figure*}
\vspace{15mm}
\epsscale{1.0}
\plotone{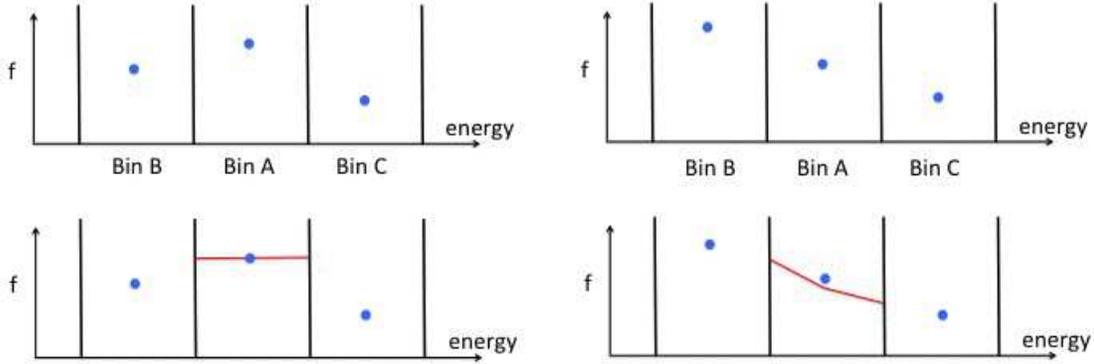}
\caption{The reconstruction of subgrid energy spectrum on {\it LRG}. The upper panels show two representative distributions of grid-point values of $f$ (blue filled-circles) for three consecutive energy bins. The lower panels present the reconstructed subgrid spectra (red lines) in Bin A for the two cases. If the grid-point value in energy bin A takes a local minimum or maximum among the three grid points (left panels), then we assume a uniform subgrid spectrum. Otherwise (right panels) we construct the subgrid spectrum iteratively. See the text for details.
\label{reconstruction_notmono}} 
\end{figure*}


In our Boltzmann solver, transformations between different energy grids are frequently performed. As mentioned in Section~\ref{sec:difficulty}, we will be able to deploy at most $\sim 20$ energy bins, a rather coarse resolution. We hence need a subgrid modeling of neutrino-energy spectrum. It is also important for computing fluxes at grid boundaries. As a matter of fact, if we did not take into account such subgrid distributions and assumed instead that neutrinos are populated uniformly in each grid, then a large number of neutrinos could artificially leak to neighboring grids either by inaccurate numerical fluxes or by imprecise interpolations (see also Step~4 on this issue).





In the reconstruction one should pay an adequate attention to the following two conditions:
\begin{itemize}
\item
monotonicity

\item
Conservation of neutrino numbers
 
\end{itemize}
The first condition is familiar in the numerical treatment of hyperbolic systems and necessary to avoid artificial generation of extrema in spectra, which may cause numerical instabilities. The importance of the second condition is rather obvious. In fact, if it were violated, neutrinos would appear or disappear just by changing energy grids. As shown later, this condition is particularly important in the evaluation of $f$ on {\it LFG}. Note that the value of $f$ on each grid point actually represents the average in the energy bin in our formulation.

The reconstruction procedures are schematically shown in Figure~\ref{reconstruction_notmono}, in which we are going to construct the subgrid energy spectrum for energy bin A in {\it LRG}. In so doing, not only grid point A but also neighboring grid points B and C are utilized. We distinguish two cases: (1) $f$ takes an extremal value locally on grid point A, i.e., both of $f$'s on grid points B and C are either larger or smaller than $f$ on grid point A; (2) otherwise.

 The left panels in Figure~\ref{reconstruction_notmono} correspond to the first case. As shown there, we assume in this case a flat spectrum in the energy bin. This is not a bad approximation, since the actual spectrum is indeed nearly flat in the vicinity of a local extremum. In the second case, in which $f$ changes monotonically over the neighboring three grid points, we reconstruct a subgrid spectrum as follows, which is shown in the right panels in Figure~\ref{reconstruction_notmono}.

 We first determine the value of $f$ on the left and right interfaces of energy bin A as the averages of adjacent grid-point values in the logarithmic scale. They are referred to as $f_{\rm{L}}$ ($f_{\rm{R}}$), respectively. We use the grid widths for the weight in the average. We also define $f_{\rm{max}}$ and $f_{\rm{min}}$ as the largest (smallest) of $f_{\rm{L}}$ and $f_{\rm{R}}$. Denoting the grid-point value of $f$ by $f_{\rm{m}}$, we first construct a trial spectrum $f_{tmp}$ as follows:
\begin{eqnarray}
&&f_{\rm{tmp}}(\varepsilon) = \frac{1}{ e^{G(\varepsilon)} + 1 } \label{eq:interpofunc} \\
&&G(\varepsilon) = 
\begin{cases}
\dfrac{G_{\rm{L}} - G_{\rm{m}}}{\varepsilon_{\rm{L}} - \varepsilon_{\rm{m}}} ( \varepsilon - \varepsilon_{\rm{m}} ) + G_{\rm{m}} & (\varepsilon < \varepsilon_{\rm{m}}), \\
\hspace{1mm} \\
\dfrac{G_{\rm{R}} - G_{\rm{m}}}{\varepsilon_{\rm{R}} - \varepsilon_{\rm{m}}} ( \varepsilon - \varepsilon_{\rm{m}} ) + G_{\rm{m}} & (\varepsilon > \varepsilon_{\rm{m}}),
\end{cases}
\end{eqnarray}
where $\varepsilon_{\rm{L}}$, $\varepsilon_{\rm{R}}$, and $\varepsilon_{\rm{m}}$ are the energies at the left and right interfaces and grid point, respectively; $G_{\rm{L}}$, $G_{\rm{R}}$ and $G_{\rm{m}}$ are the corresponding values of $G$ given as
\begin{eqnarray}
G_{\rm{i}} \equiv {\rm log}( \frac{1}{f_{\rm{i}} - 1}) \hspace{3mm} (\rm{i}= \rm{L, R, m}). \label{eq:Gi}
\end{eqnarray}
It is clear that this expression becomes exactly correct if neutrinos are in thermal equilibrium and take a Fermi-Dirac distribution.

 It is obvious, however, that $f_{\rm{tmp}}(\varepsilon)$ does not ensure the conservation of neutrino number. We hence need to modify $f_{\rm{tmp}}(\varepsilon)$. We first integrate $f_{\rm{tmp}}$ in the energy bin to obtain the neutrino number, $N^{'}_{\rm{A}}$ in it. This should have been equal to $N_{\rm{A}}$, the true value. Using the ratio,
\begin{eqnarray}
R_{\rm{rate}} \equiv \frac{N_{A}}{N^{'}_{A}}, \label{eq:numberratio}
\end{eqnarray}
we scale the temporary spectrum as $f_{\rm{tmp}} \times R_{\rm{rate}}$ to obtain a new subgrid spectrum, which by definition satisfies the conservation of neutrino number exactly.

The new spectrum so obtained do not satisfy the monotonicity condition in general, which requires that $f$ should always lie between $f_{\rm{max}}$ and $f_{\rm{min}}$. We hence apply a limiter if $f_{\rm{tmp}}$ exceeds $f_{\rm{max}}$ and/or $f_{\rm{min}}$: $f_{\rm{tmp}}$ are modified so that they would lie in between. Owing to this limiter, $N^{'}_{\rm{A}}$ obtained by integration of the new subgrid spectrum, again deviates from $N_{\rm{A}}$. We repeat the above procedure until the following condition is satisfied:
\begin{eqnarray}
|1 - R_{\rm{rate}}| < \epsilon_{\rm{conv}}, \label{eq:convergencecondi}
\end{eqnarray}
where $\epsilon_{\rm{conv}}$ is a measure of convergence and is set to $\epsilon_{\rm{conv}}=10^{-3}$. It is important that the convergence of this iteration is guaranteed mathematically and that no artificial extremum emerges in the reconstructed spectrum.

\subsection{Step~3: Lagrangian Remapping} \label{subsec:step3}

\begin{figure*}
\vspace{15mm}
\epsscale{1.0}
\plotone{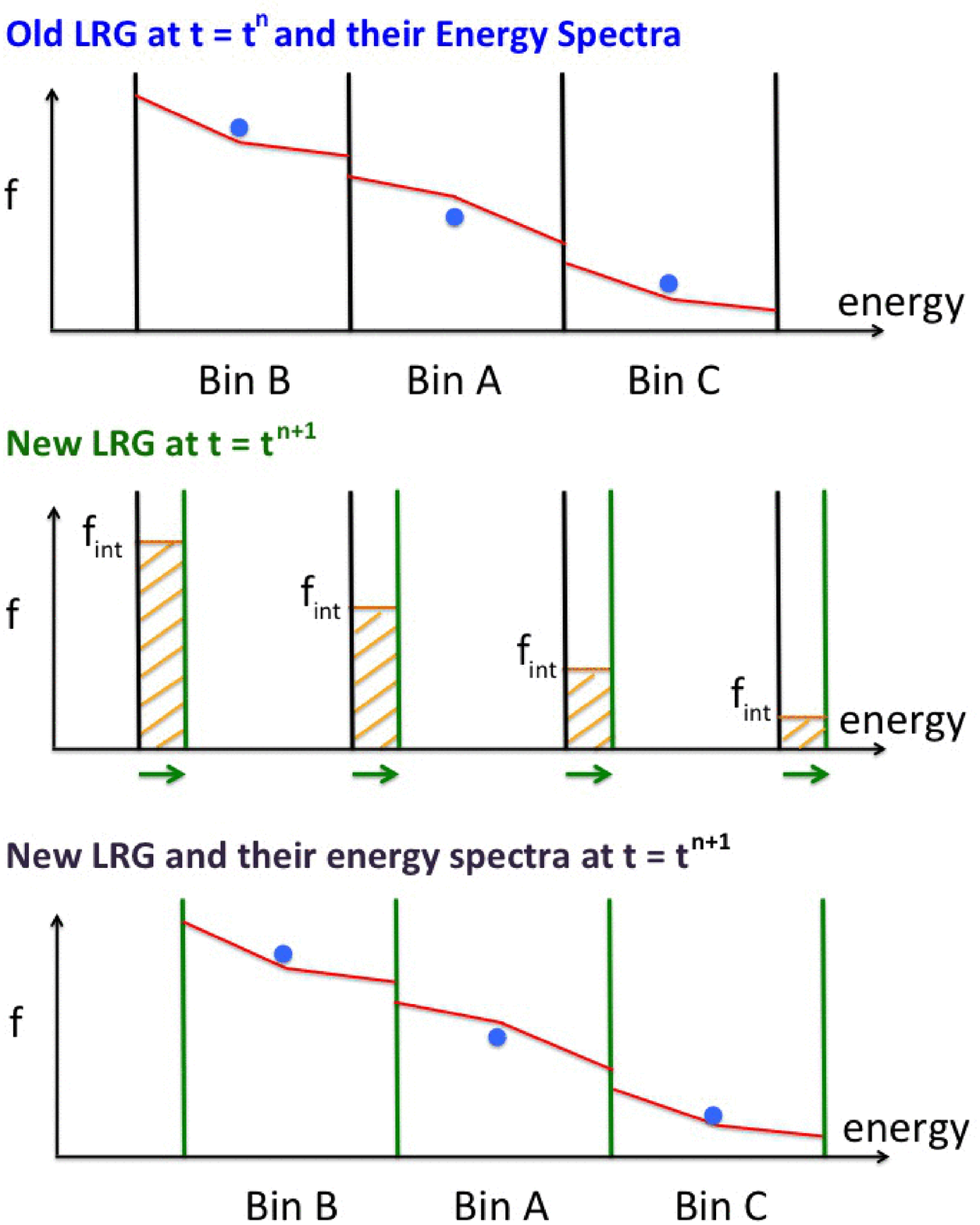}
\caption{The Lagrangian remapping. The upper panel shows the grid-center values of $f$ and the reconstructed subgrid energy spectra on the {\it LRG} at $t=t^{n}$. The middle panel compares the {\it LRG}'s at $t=t^{n}$ (black lines) and $t=t^{n+1}$ (green lines). Arrows indicate the shifts of grid interfaces. The neutrinos in shaded regions should change affiliations to the neighboring grids (see in the text for more details). The lower panel displays the grid-center values of $f$ and the reconstructed subgrid spectra on the new {\it LRG} at $t=t^{n+1}$.
\label{eneshiftfreconst}} 
\end{figure*}

 Here we carry out the Lagrangian remapping of neutrino-energy grids and compute the change in $f$ on {\it LRG}. The subgrid energy spectrum, which is obtained in the previous step, play an important in this process.

The procedure is summarized in Figure~\ref{eneshiftfreconst}. Suppose that $n$ time integrations have been finished and all quantities associated with matter and neutrinos have been obtained at $t = t^{n}$. The upper panel shows the grid-point values of $f$ as well as subgrid spectra in three consecutive energy binges on {\it LRG} at $t=t^{n}$. Note that the angular dimensions are suppressed in the figure. In Step~1, matter velocities are changed. As explained in Section~\ref{sec:twoenegrids}, {\it LRG} is determined so that the neutrino-energy grid be identical in each instantaneous fluid-rest frame. We hence need to update {\it LRG} accordingly as shown with green lines in the middle panel of Figure~\ref{eneshiftfreconst}. It is then evident that $f$ should be also changed on account of the shifts of the grid boundaries. As shown in the figure, neutrinos in the shaded areas determine the change of $f$ due to this remapping.

To evaluate the numbers of neutrinos in these regions, we use $f_{\rm{int}}$, the interface value of $f$ on the old {\it LRG} at $t=t^{n}$. It is obtained as the smaller one of the two interface values derived from the subgrid spectra in the adjacent grids in order to prevent the moving of too many neutrinos. With this $f_{\rm{int}}$ and the energies at the grid interfaces on the old ($\varepsilon_{\rm{old}}$) and new ($\varepsilon_{\rm{new}}$) {\it LRG}'s, the number of neutrinos to be moved to the adjacent grid $\Delta N_{\nu}$ is given by
\begin{eqnarray}
\Delta N_{\nu} = f_{\rm{int}} \frac{1}{3}
| (\varepsilon_{\rm{new}})^3 - (\varepsilon_{\rm{old}})^3 | \Delta \Omega ,
 \label{eq:neutrinonumshift}
\end{eqnarray}
where $\Delta \Omega$ is the extent of solid angle of the bin\footnote{Note again that we suppress the angular dimension in Figure~\ref{eneshiftfreconst}.}.
Note that the conservation of neutrino number is automatically guarantied in this process. We end up this step with a construction of the subgrid spectrum for the modified neutrino number (and accordingly $f$) in each energy bin just in the same way as in Step~2.

It is emphasized that no interactions of neutrinos with matter have been taken into account yet up to this point. The change in $f$ considered above is induced by the acceleration or deceleration of matter (and hence of the local fluid-rest frame). As explained in Section~\ref{sec:twoenegrids} in detail, if such a change in matter velocity occurred in optically thin matter, the neutrino energy spectrum should not change in the laboratory frame. The red or blue shifts of the energy spectrum in the fluid-rest frame are compensated for by the Lagrangian remapping in our method. It should be also noted that the energy shift is proportional to the time step $\Delta t$ and is small as a consequence, the fact which is true even in the shock wave. This not only justifies the above estimation of $\Delta N_{\nu}$ but also is a huge advantage in the use of {\it LRG} compared with other Lagrangian formulations, in which large energy shifts may occur.




\subsection{Step~4: Evaluations of Advection and Collision Terms} \label{subsec:step4}

\begin{figure*}
\vspace{15mm}
\epsscale{1.2}
\plotone{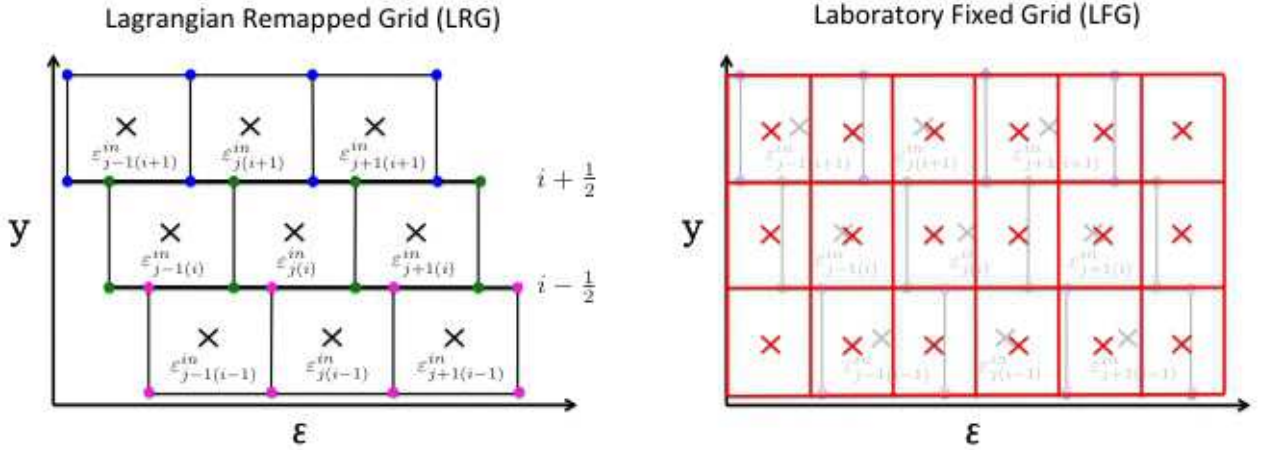}
\caption{Left: Energy bins in {\it LRG} at neighboring spatial or angular points. $y$ denotes a spatial or angular dimension, whose grid points are specified by the subscript, $\rm{i}$. The subscript $j$ indicates the energy grid points. Right: The same as the left panel but in LFGs (red rectangules). The energy bins and grid points in {\it LRG} are also displayed in gray for comparison in this panel.
\label{LRG_LFGgrids}}
\end{figure*}

\begin{figure*}
\vspace{15mm}
\epsscale{1.2}
\plotone{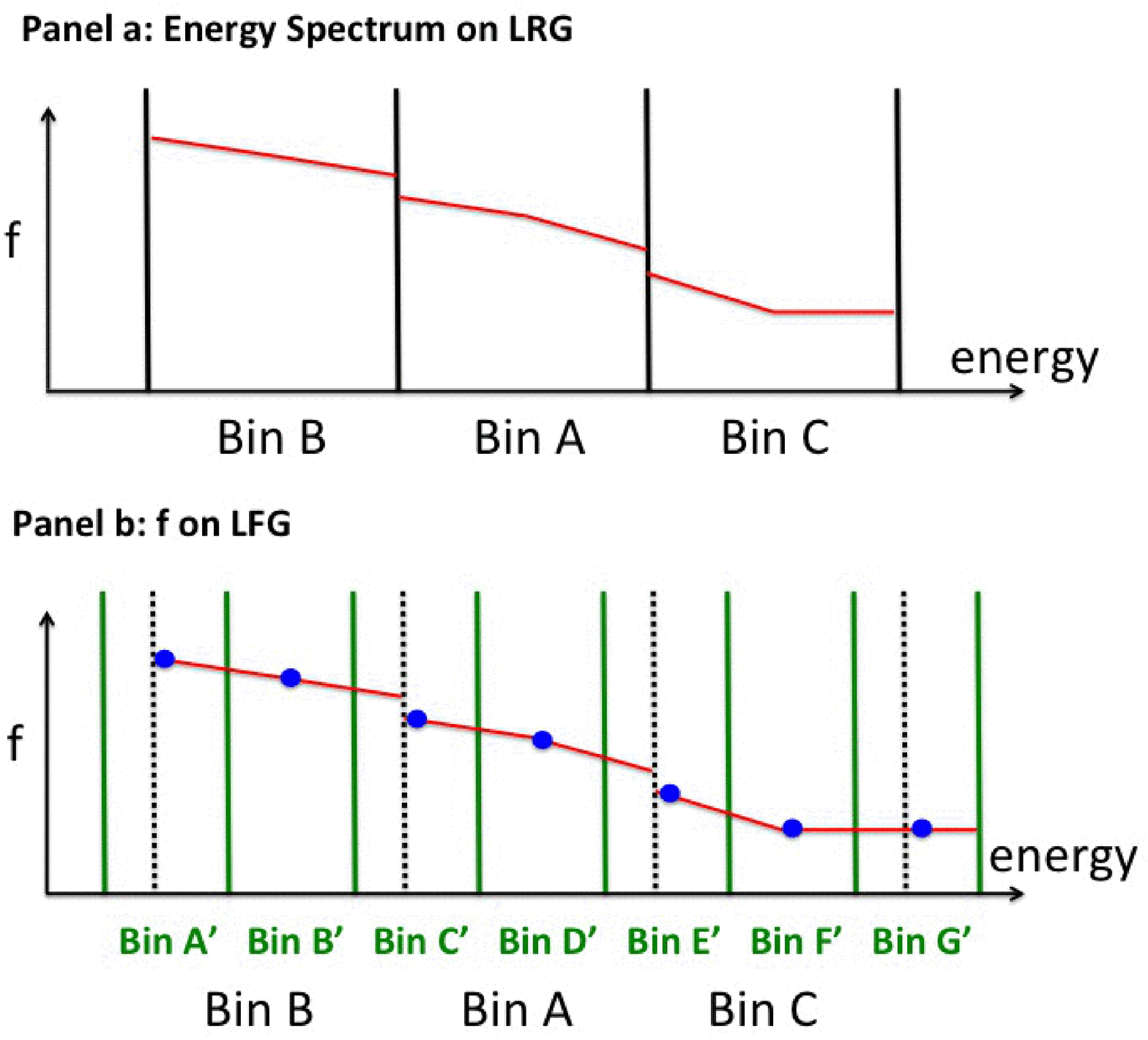}
\caption{Upper panel: The subgrid energy spectrum on {\it LRG} at $t=t^{n}$. Lower panel: The same as the upper panel but on {\it LFG} (shown in green). For comparison, {\it LRG} is displayed by dotted lines. The blue filled circles represent the grid-point values of $f$ on {\it LFG}.
\label{LFgridfigure}} 
\end{figure*}

Now that the energy shift induced by matter motions has been treated, the remaining task is to consider the spatial advections and collisions of neutrinos. The latter is easy on {\it LRG}, which is essentially the comoving grid, and will be briefly explained at the end of this section. The former, on the other hand, is fairly complicated, since {\it LRG} is not uniform in space. In contrast to the ordinary Lagrangian formulation, in which the spatial advection is expressed as complicated partial derivatives, our method utilizes the fact that the advection is very simple in the laboratory frame. {\it LFG} defined in Section~\ref{sec:twoenegrids} is the main tool here.


As explained in Section~\ref{sec:twoenegrids}, {\it LFG} is defined so that the energy grids are identical everywhere in space and it does not depend on the flight direction of neutrino. In addition {\it LFG} should have the following properties:
\begin{itemize}
\item
{\it LFG} covers the union of all energy ranges in {\it LRG} \footnote{Note that the energy range covered by {\it LRG} depends on the spatial position and flight direction.}.

\item
Each energy bin in {\it LRG} is covered by more than one energy bins in {\it LFG}. 
\end{itemize}
These conditions are important to ensure the accuracy in the evaluation of the advection terms. Figure~\ref{LRG_LFGgrids} displays the example of the relation between the two grids in the laboratory frame.


Given {\it LFG}, we evaluate the advection term as follows. We suppose that the n-th time step has been completed and $f^{n}$ is given on each {\it LRG} and the subgrid spectrum has also been constructed according to Step~2 (see panel~a in Figure~\ref{LFgridfigure}). Using this subgrid spectrum, which is denoted by $f_{\rm{sub}}$ in this section, we assign $f$ to each grid point in {\it LFG}, which is denoted as $f_{\rm{LF}}$:
\begin{eqnarray}
f_{\rm{LF}} \equiv f_{\rm{spe}} (\varepsilon_{\rm{LFm}}), \label{eq:fLF}
\end{eqnarray}
where $\varepsilon_{\rm{LFm}}$ corresponds to the neutrino energy at the grid point in {\it LFG} (see panel~b in the same figure, in which {\it LFG} is presented in green while {\it LRG} is shown with black dots.). Here we would like to emphasize again that the spatial and angular advection\footnote{Hereafter the angular advections mean the advection of neutrinos in the two-dimensional momentum subspace spanning all the flight directions.} terms on {\it LFG} (the left hand side of Eq.~(\ref{eqn:eqtransfin-spherical})) are very simple. In fact, we can employ the same method as used in \citet{2012ApJS..199...17S}\footnote{In this method the upwind and central finite-differences are interpolated according to the optical depth. In so doing, we introduce the weighting factor, $\beta$, which is linearly interpolated from {\it LRG} to {\it LFG} in the present formulation. Since $\beta$ takes a value in the range of 0.5 to 1 and does not strongly depend on the neutrino energy unlike $f$, the simple linear interpolation is justified.}.


Once the numerical fluxes for the spatial and angular advections are obtained on {\it LFG}, we then calculate the corresponding numerical fluxes for {\it LRG} as follows. We take as an example panel b in Figure~\ref{LFgridfigure}. As mentioned earlier, {\it LFG} is finer than {\it LRG}. Energy bin A in {\it LRG}, for instance, is covered by three bins C', D' and E' in {\it LFG}. Let us look at bin E' in {\it LFG}, which overlaps with bins A and C in {\it LRG}. The numerical flux in {\it LFG} should be hence shared with the latter two bins in {\it LRG}. For that purpose, we introduce a factor, $\gamma$, as defined shortly and divide the numerical flux $F_{E'}$ into $\gamma F_{E'}$ and $(1-\gamma) F_{E'}$. $\gamma$ is defined as
\begin{eqnarray}
\gamma \equiv \frac{N_{\rm{L}}}{N_{\rm{L}} + N_{\rm{R}}},
\end{eqnarray}
with
\begin{eqnarray}
N_{\rm{L(R)}} = |( \varepsilon_{\rm{AC}}^3 - \varepsilon_{\rm{L(R)}}^3)| f_{\rm{A(B)}}, \label{eq:distributionrate}
\end{eqnarray}
where $\varepsilon_{\rm{AC}}$ is the neutrino energy at the interface of bins A and C in {\it LRG} and $\varepsilon_{\rm{L(R)}}$ is the energy at the left (right) boundary of bin E' in {\it LFG}; $f_{\rm{A(B)}}$ is the grid-point value of $f$ for bin A(B) in {\it LRG}. The numerical flux for bin A of {\it LRG} is a sum of the contributions from bins C', D' and E' in {\it LFG}, each obtained in this manner.

So far we have explained our treatment of the spatial and angular advection terms as if they were finite-differenced explicitly in time. As a matter of fact, we treat them implicitly in our method. This is important from a point of view of numerical stability and computational times. In the following, the detailed procedure is described.

We first rewrite the Boltzmann equation (Eq.~(\ref{eqn:eqtransfin-spherical})) as
\begin{eqnarray}
\frac{\partial f}{\partial t} + \sum_{k} F^{k}_{\rm{adv}}(f) = \Bigl( \frac{\delta f}{\delta t} \Bigr)_{{\rm col}}^{{\rm lb}}(f), \label{eq:conBoltzrewrite}
\end{eqnarray}
in which the second term on the left hand side is the sum of spatial and angular advection terms. Since the following treatment is common to each term in the sum, we drop the subscript $k$ hereafter.

In the implicit approach, both the advection and collision terms are evaluated at $t=t^{n+1}$ and the finite-differenced equation is written as
\begin{eqnarray}
\frac{f^{n+1} - f^{n}}{\Delta t} = - F_{\rm{adv}}(f^{n+1}) + \Bigl( \frac{\delta f}{\delta t} \Bigr)_{{\rm col}}^{{\rm lb}}(f^{n+1}). \label{eq:conBoltzrewrite_fullimp}
\end{eqnarray}
It is noted, however, that $F_{\rm{adv}}$ in SR is evaluated via the complex interpolation of $f$ between {\it LFG} and {\it LRG} and is nonlinear and highly complicated. This prevents us even from linearizing the equation\footnote{But the collision terms can be easily treated implicitly. See the discussion of the end of this section.} and the implicit treatment of advection terms seems impossible.

It is noteworthy, however, that in the Newtonian approximation, in which no distinction is made between the fluid-rest and laboratory frames, the advection term is linear and can be treated completely implicitly. In fact the resultant non-relativistic (NR) equation can be cast into the following form (see also \citet{2012ApJS..199...17S}):
\begin{eqnarray}
F^{\rm{NR}}_{\rm{adv}}(f^{n+1}) = A f^{n+1}_{(i)} + B f^{n+1}_{(i+1)} + C f^{n+1}_{(i-1)}, \label{eq:nonrelaadv}
\end{eqnarray}
where $F^{\rm{NR}}_{\rm{adv}}$ is the NR advection term and the subscript $i$ indicates the spatial and angular grid points symbolically. Coefficients $A, B$, and $C$ are written as a function of space and flight directions. This fact suggests the following scheme for the advection term in SR:
\begin{eqnarray}
&&F_{\rm{adv}} = F^{SR}_{\rm{adv}}(f^{n}) + 
\Bigl(    
F^{\rm{NR}}_{\rm{adv}}(f^{n+1}) - F^{\rm{NR}}_{\rm{adv}}(f^{n})
 \Bigr). \label{eq:stabilizationA}
\end{eqnarray}
The first term on the right hand side is the relativistic advection term evaluated with $f^{n}$ and the second term in the parenthesis is a correction term. Eq.~(\ref{eq:stabilizationA}) is only semi-implicit as it is. We hence replace $f^{n}$ in Eq.~(\ref{eq:stabilizationA}) by a trial value, $f^{\rm{gs}}$ and repeatedly solve the Boltzmann equation (Eq.~(\ref{eq:conBoltzrewrite})), updating $f^{\rm{gs}}$ with $f^{n+1}$ obtained for $f^{\rm{gs}}$ until they coincide with each other within a certain error. This ensures full-implicitness of our method as explained shortly.


The idea behind this method should be clear. If matter motion is not very fast compared with the speed of light, which is indeed the case in CCSNe, $F^{SR}_{\rm{adv}}(f^{\rm{gs}})$ is almost equal to $F^{\rm{NR}}_{\rm{adv}}(f^{\rm{gs}})$, and $F_{\rm{adv}}$ will be dominated by $F^{\rm{NR}}_{\rm{adv}}(f^{n+1})$. Then this scheme is close to the NR implicit scheme. In addition, there are other important properties in the prescription: firstly if $f^{\rm{gs}}$ coincides with $f^{n+1}$, as is the case at the end of iterations, the correction term vanishes and only $F^{SR}_{\rm{adv}}(f^{\rm{gs}}) (= F^{SR}_{\rm{adv}}(f^{n+1}))$ remains. This property guarantees the full implicitness of our scheme. It is also mentioned that $\Delta t$ is actually limited most of times by the requirement that $f$ should not change by more than a few percent in a single time step, and the correction term is a small correction to the first term in most situations (but see below for the exceptional case.). In spite of this, we find that this correction term significantly improves the numerical stability, enabling us to take larger $\Delta t$.

%

Although the above method works fairly well most of times, it fails sometimes when the matter velocity reaches several tens percent of the speed of light and the correction term becomes significantly larger than $F^{SR}_{\rm{adv}}(f^{\rm{gs}})$. In such cases the iteration does not converge unless we reduce $\Delta t$. To avoid too small a value of $\Delta t$, however, we introduce a limiter $\kappa$
\begin{eqnarray}
&&F_{\rm{adv}} = F^{SR}_{\rm{adv}}(f^{\rm{gs}}) + 
\kappa \Bigl(    
F^{\rm{NR}}_{\rm{adv}}(f^{n+1}) - F^{\rm{NR}}_{\rm{adv}}(f^{\rm{gs}})
 \Bigr). \label{eq:stabilizationB}
\end{eqnarray}
in which $\kappa$ is determined so that the correction term should not exceed the first term. It is stressed that such a prescription is just a technique to improve the convergence and does not affect the final result, since the second term vanishes in the end anyway.


We turn to the collision terms before closing this section. There are no new difficulties in their treatment, since {\it LRG} is essentially the same as the fluid-rest frame employed in the ordinary Lagrangian methods. There is one feature, however, which is original in our method.

The angular dimensions in momentum space are discretized in the same manner everywhere on {\it LRG} (see the left panel of Figure~\ref{momspacedist_laremap}). This means that the angular gridding is not uniform in the fluid-rest frame due to aberration by Lorentz transformation. The angular integration as shown in Eq.~(\ref{eq:isoenergyscatkenel}) for the isoenergetic scattering is normally performed in the fluid-rest frame. In our approach, however, that is done on {\it LRG} in the laboratory frame. In so doing, the aberration effect is taken account of as the Jacobian as follows
\begin{eqnarray}
\int A(\Omega^{\rm{fr}}) d \Omega^{\rm{fr}} = \int B(\Omega^{{\rm lb}}) \Bigl( \frac{d \Omega^{\rm{fr}} }{d \Omega^{{\rm lb}} } \Bigr)  d \Omega^{{\rm lb}}, \label{eq:omegainteg}
\end{eqnarray}
where $A$ is an arbitrary function of $\Omega^{\rm{fr}}$ and $B$ is defined as $B(\Omega^{{\rm lb}}) \equiv A(\Omega^{\rm{fr}}(\Omega^{{\rm lb}}))$. The Jacobian ($d \Omega^{\rm{fr}} / d \Omega^{{\rm lb}} $) has already been derived in Eq.~(\ref{eqn:solidrela}).

As mentioned earlier, the collision terms are treated fully implicitly. The point is that the matrix structure originated from the collision terms is exactly the same as the one for the NR case, which implies that the numerical tools developed for our Newtonian code \citep{2012ApJS..199...17S} can be utilized for the present code as they are. As a matter of fact, thanks to this implicit treatment of collision terms the time steps $\Delta t$ in the 1D test simulation of CCSNe (see Section~\ref{subsec:1DSRBoltz_Hydro}) are comparable to those in our previous simulations with a 1D implicit GR Lagrangian Boltzmann-Hydro code \citep{2005ApJ...629..922S}.

\subsection{Step~5: Feedbacks to matter} \label{subsec:step5}
Solving the Boltzmann equations in the previous step, we now treat feedbacks from the neutrino-matter interactions to hydrodynamics. The hydrodynamics equations and the conservation equation for electron number are written as\footnote{See Eqs.~(\ref{eq:hydroConservativeform})--(\ref{eq:systemeqcompo})}
\begin{eqnarray}
T_{\rm{hd} \hspace{1.5mm} ,\nu}^{\mu \nu} &=& - G^{\mu}, \label{eq:TandGfinal} \\
N_{e \hspace{1mm} ,\nu}^{\nu} &=& - \Gamma, \label{eq:NandGammafinal}
\end{eqnarray}
where the right hand sides correspond to the feedbacks and are written as
\begin{eqnarray}
G^{\mu} &\equiv& \sum_{\rm{i}} G_{\rm{i}}^{\mu}, \label{eq:Gsumdef} \\
G_{\rm{i}}^{\mu} &\equiv& \int p_{\rm{i}}^{\mu} \Bigl( \frac{\delta f}{\delta \tau} \Bigr)_{\rm{col}(\rm{i})} dV_p, \label{eq:Gdef} \\
\Gamma &\equiv& \Gamma_{\nu_{e}} - \Gamma_{\bar{\nu_{e}}}, \label{eq:Gammasumdef} \\
\Gamma_{i} &\equiv& \int \Bigl( \frac{\delta f}{\delta \tau} \Bigr)_{\rm{col}(\rm{i})} dV_p. \label{eq:Gammadef}
\end{eqnarray}
In these expressions, the invariant volume in the momentum space is denoted by $dV_{p}$ and the subscript "$\rm{i}$" indicates the neutrino species.

At the very end of all steps, we again perform Steps~2 and 3, since matter velocities are changed due to the momentum exchange between matter and neutrinos. This closes the update from $t=t^{n}$ to $t=t^{n+1}$. We iterate these steps as many times as needed.



\section{Validation} \label{sec:validation}
In order to validate our new formulation of SR Boltzmann Radiation-Hydrodynamics, we carry out a series of code tests. We first focus on the Boltzmann solver, i.e., the feedbacks to hydrodynamics are ignored. We test the advections and collisions separately in idealized setups in order to see the code performance in each sector clearly. In these tests, only electron-type neutrinos are taken into account, since the treatments of SR effects are common to other species.

We then perform SR Boltzmann-Hydro simulations of 1D spherical core collapse for the $15 M_{\sun}$ progenitor. In these test runs, we consider 3 species of neutrinos ($\nu_{e}$, $\bar{\nu_{e}}$, and $\nu_{x}$) and implement minimal but essential microphysics. For comparison, a NR simulation is also performed. Based on the two results, we discuss the importance of SR effects.

\subsection{Collision term: isoenergetic scattering} \label{subsec:colisoenesca}

\begin{figure*}
\vspace{15mm}
\epsscale{0.8}
\plotone{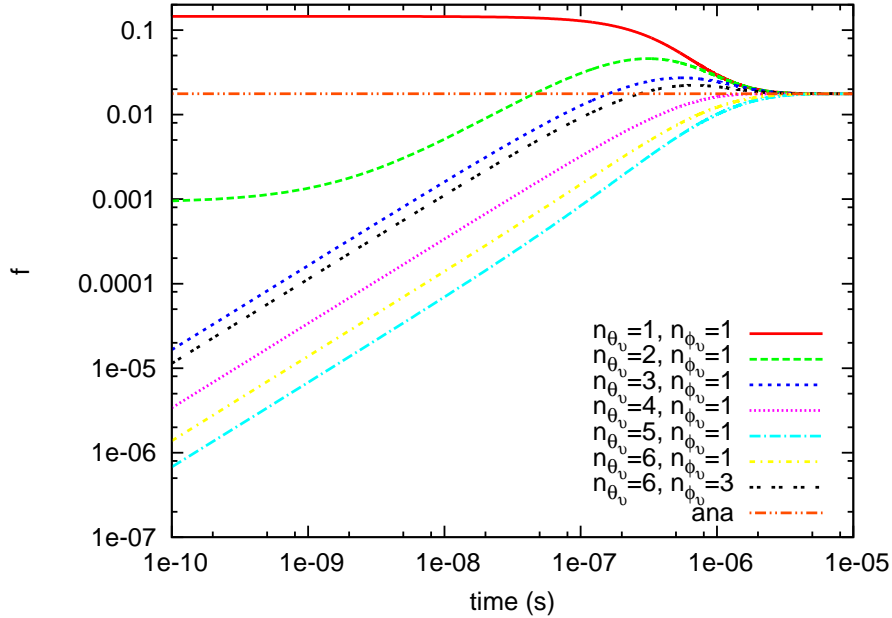}
\caption{The time evolutions of $f$ for different angles by the isoenergetic scatterings. $n_{\theta_{\nu}}$ and $n_{\phi_{\nu}}$ specify the angular grid points. The neutrino energy is set to $60$~MeV in the fluid-rest frame. The orange dash-dotted line indicates the final state of $f$ at $t \rightarrow \infty$ obtained analytically.
\label{scattest1}} 
\end{figure*}

\begin{figure*}
\vspace{15mm}
\epsscale{1.0}
\plotone{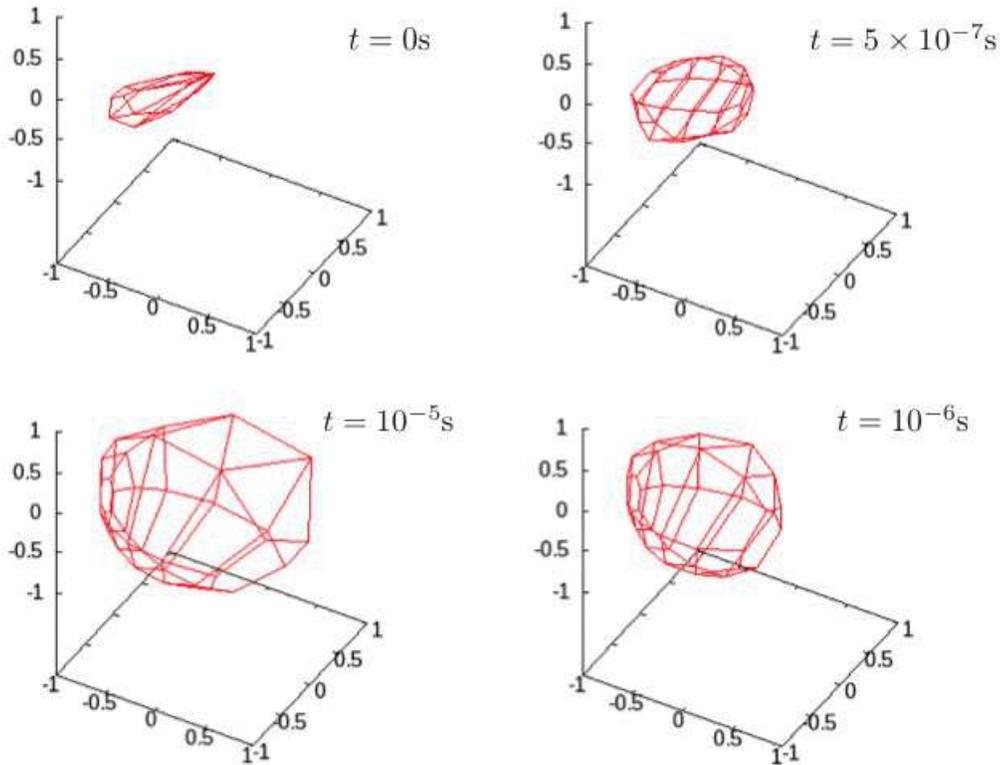}
\caption{3D presentations of $f$ in the fluid-rest frame at different times. Only the angular distributions are shown for neutrinos with an energy of $60$~MeV in the fluid-rest frame. See the text for more details on the construction of wire frames.
\label{scattest2}} 
\end{figure*}

\begin{figure*}
\vspace{15mm}
\epsscale{1.0}
\plotone{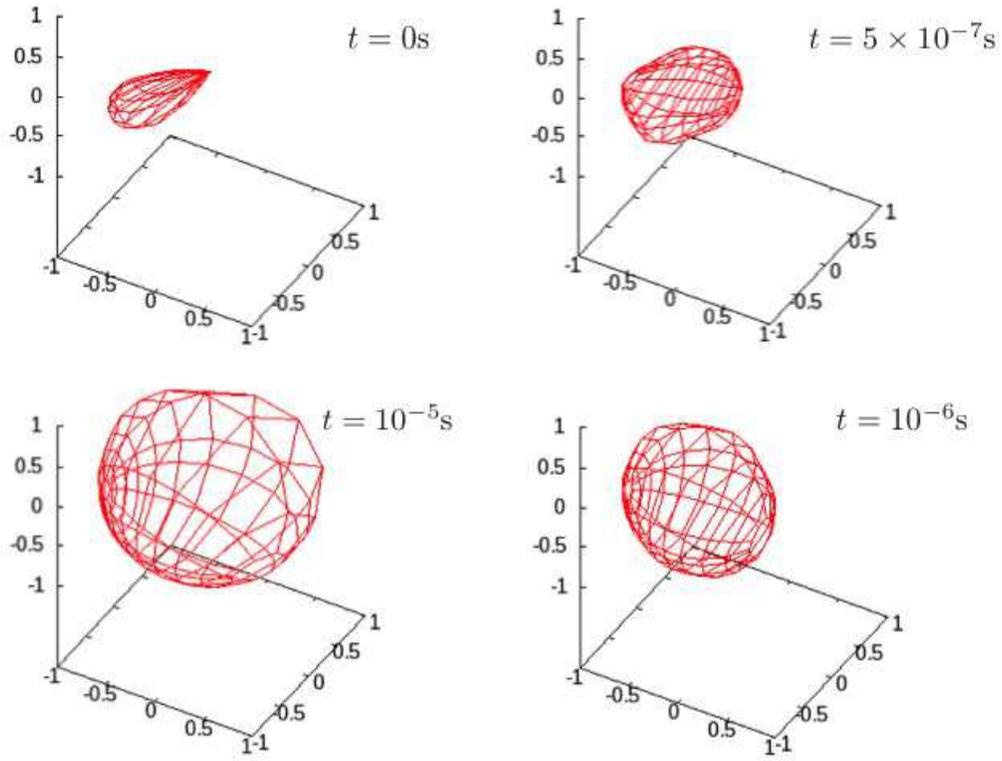}
\caption{The same as Figure~\ref{scattest2} but for a higher angular resolution ($N_{\theta_{\nu}}=12$, $N_{\phi_{\nu}}=12$).
\label{scattest3}} 
\end{figure*}

\begin{figure*}
\vspace{15mm}
\epsscale{0.5}
\plotone{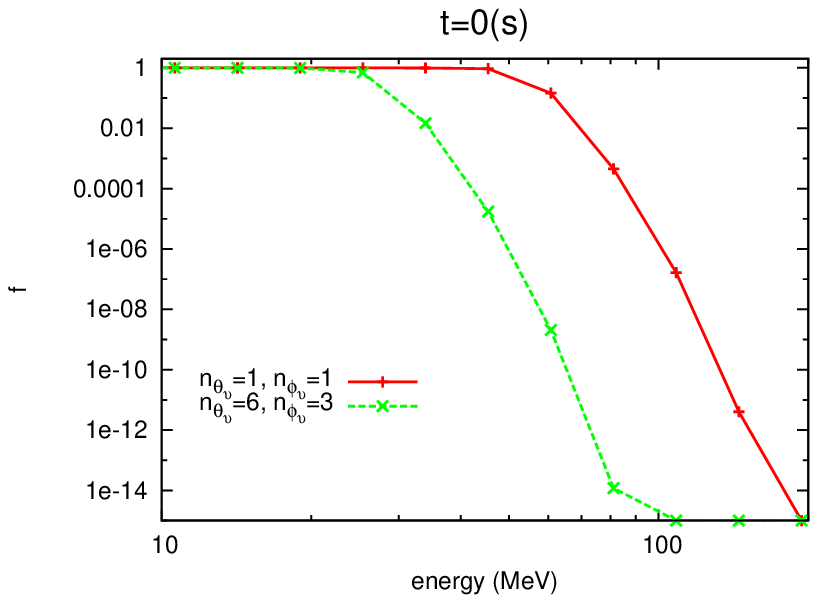}
\plotone{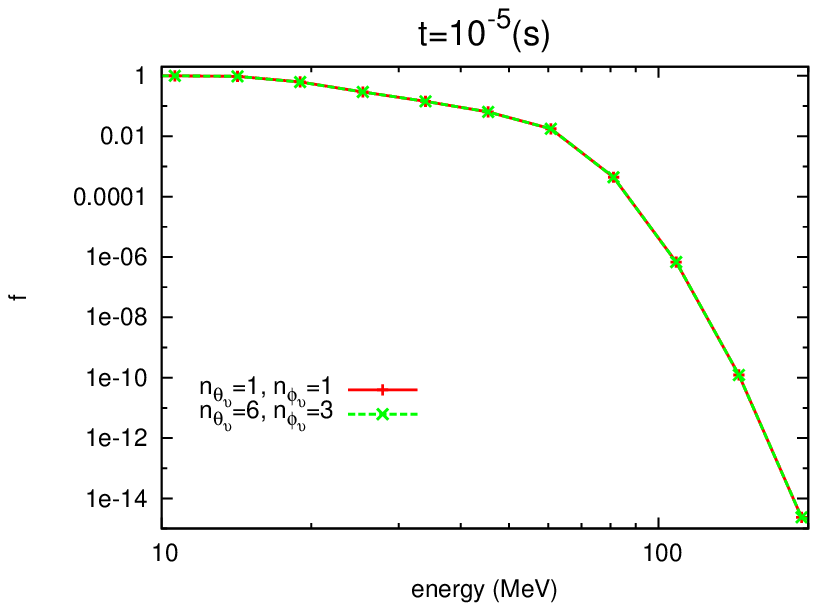}
\caption{The energy spectra for different flight directions in the fluid-rest frame at two different times (left: $t=0$~s, right: $t=10^{-5}$~s). Note that the floor value of $f$ is set as $10^{-15}$, which is observed at high energies.
\label{scattest4}} 
\end{figure*}

\begin{figure*}
\vspace{15mm}
\epsscale{0.5}
\plotone{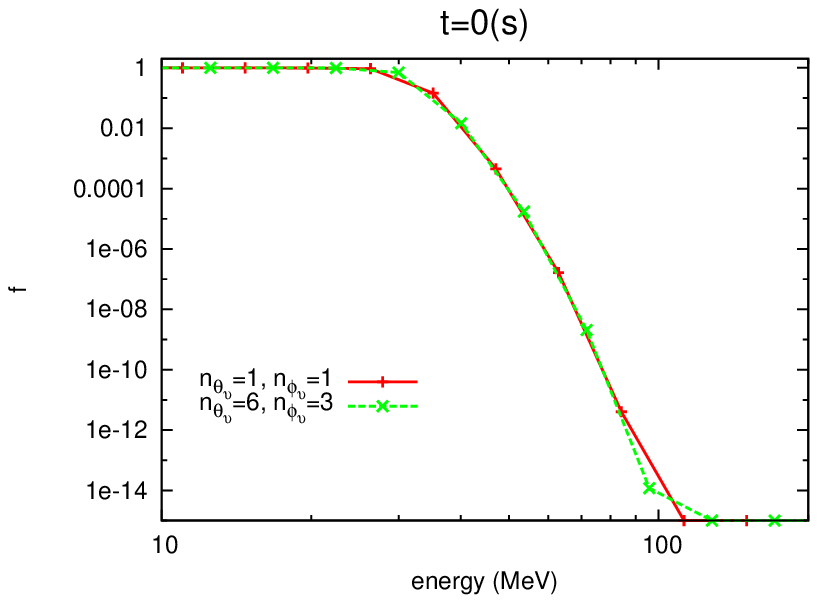}
\plotone{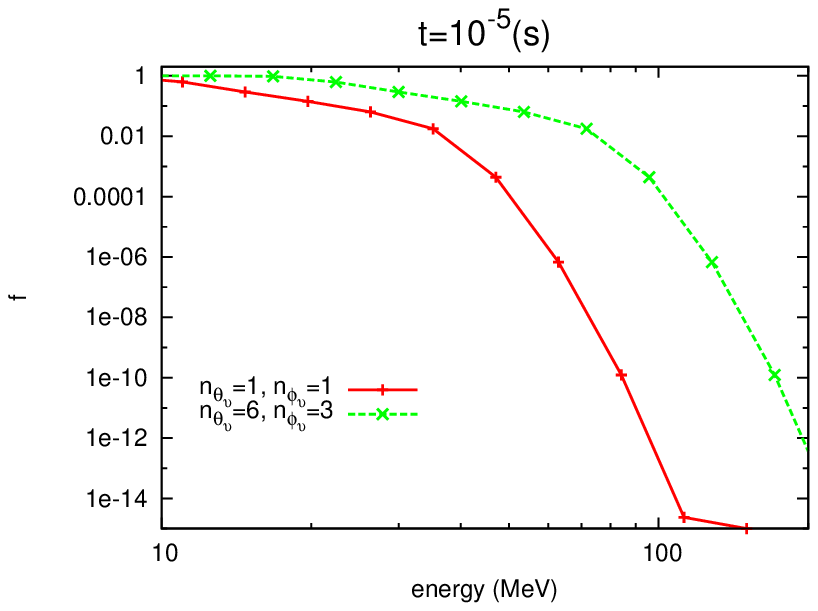}
\caption{The same as Figure~\ref{scattest4} but in the laboratory frame.
\label{scattest5}} 
\end{figure*}

As discussed in Section~\ref{sec:difficulty}, the isoenergetic scattering is the primary source of difficulties in the $S_{n}$ method for the SR Boltzmann equation. This test is meant to see whether our code can properly handle this process. This is a single zone calculation, in which we deploy only one spatial grid and the advection term is neglected. We are concerned only with the collision term. Hydrodynamical quantities are assumed to be constant in time and set as $\rho=10^{12} {\rm g}/{\rm cm}^3$, $T = 2$MeV, and $Y_{e} = 0.4$, where $\rho$, $T$, and $Y_e$ denote the density, temperature and electron fraction, respectively. Under this thermodynamical condition, free nucleons and nuclei are both existent. We hence consider the following isoenergetic scatterings:
\begin{eqnarray}
\nu + N &\longleftrightarrow& \nu + N \hspace{2mm} , \\
\nu + A &\longleftrightarrow& \nu + A \hspace{2mm} .
\end{eqnarray}

Although we drop the advection term in this test, we set a non-vanishing velocity as follows:
\begin{eqnarray}
v^{r} &=& v \hspace{1mm} {\rm cos}~\theta_{h}, \\
v^{\theta} &=& v \hspace{1mm} {\rm sin}~\theta_{h} \hspace{1mm} {\rm cos}~\phi_{h}, \\
v^{\phi} &=& v \hspace{1mm} {\rm sin}~\theta_{h} \hspace{1mm} {\rm sin}~\phi_{h}, \label{eq:vdistri_forcol}
\end{eqnarray}
where $v^{r}$, $v^{\theta}$, and $v^{\phi}$ denote the radial, $\theta$--, and $\phi$-- components, respectively. They are assumed to be constant in time and are controlled by three parameters, $v$, $\theta_{h}$ and $\phi_{h}$. In this test, we set $v=2 \times 10^{10} {\rm cm}/{\rm s}$, $\theta_{h}=\pi/4$, and $\phi_{h}=\pi/4$, respectively. Note that this velocity is considerably large by the CCSNe standard.

We assume that neutrinos are distributed isotropically in the laboratory frame initially, and they have Fermi-Dirac distributions in energy. The neutrino chemical potential can be obtained by the assumption that neutrinos are chemical equilibrium with matter. Since matter has a non-vanishing velocity, neutrinos are initially {\it anistropic} in the fluid-rest frame. Then, $f$ should evolve towards an isotropic distribution in the latter frame by the isoenergetic scattering.

For this test, momentum space is covered with a grid of $N_{\epsilon}(=20)$ points in energy and $N_{\theta_{\nu}}(=6) \times N_{\phi_{\nu}}(=6)$ points in angles. The gridding of {\it LRG} has been explained in detail in Section~\ref{sec:twoenegrids} and Figure~\ref{momspacedist_laremap} (see also \citet{2012ApJS..199...17S} for the construction of angular grid).

We show the numerical results in Figures~\ref{scattest1} to \ref{scattest5}. Figure~\ref{scattest1} displays the evolutions of $f$ for different angles but with the same energy ($\varepsilon^{\rm{fr}}=60$~MeV) in the fluid-rest frame. As is expected, initially different values of $f$ are changed by the isoenergetic scatterings and converge to a certain value by the time $t \sim 10^{-5}$s. Note that we work on {\it LRG} in the laboratory frame and these results are obtained by the Lorentz transformation to the fluid-rest frame. The final isotropic distribution, $f_{\rm{iso}}$, can be obtained analytically from the initial condition, $f_{\rm{ini}}$, since isoenergetic scatterings do not change the number of neutrinos:
\begin{eqnarray}
f_{\rm{iso}}(\varepsilon^{\rm{fr}}) = \frac{
{\displaystyle \sum_{\rm{i}}} \hspace{1mm} f_{\rm{ini}}(\varepsilon^{{\rm lb}}(\varepsilon^{\rm{fr}}),\Omega^{{\rm lb}(\rm{i})}) \hspace{1mm} (D^{{\rm lb}(\rm{i})})^{-3} \hspace{1mm} (\Delta \Omega^{{\rm lb}(\rm{i})})
}
{
{\displaystyle \sum_{\rm{i}}} \hspace{1mm} (D^{{\rm lb}(\rm{i})})^{-3} \hspace{1mm} (\Delta \Omega^{{\rm lb}(\rm{i})})
},
\label{eq:isotropicf}
\end{eqnarray}
where the subscript $\rm{i}$ specifies the angular grid points. It is evident from the figure that the correct results are obtained numerically.

Figure~\ref{scattest2} shows the angular evolution of $f$ on an iso-energy surface with $\varepsilon^{\rm{fr}}=60$~MeV in the fluid-rest frame. In the figure, wire-framed pictures are drawn as follows: for each angular grid point, a node is placed at a distance proportional to the value of $f$ in the corresponding direction; the neighboring nodes are then connected by lines; we use the normalization, in which the maximum distance should be unity. As a consequence, an isotropic distribution corresponds to the unit sphere in this figure.

 At the beginning (top left panel), the wire frame is elongated in one direction, indicating that the angular distribution is highly anisotropic. As time passes, however, it changes shapes and eventually ($t \sim 10^{-5}$s) becomes isotropic although it may not appear so. This is due to the rather low resolution in this computation. Indeed, Figure~\ref{scattest3}, which presents the result for a higher resolution ($N_{\theta_{\nu}}=12$, $N_{\phi_{\nu}}=12$), more clearly isotropy of the final distribution. It is also reminded that the angular grid is not uniform in the fluid-rest frame due to aberration (it is uniform in the laboratory frame. See Section~\ref{sec:twoenegrids}).

As an alternative presentation of isotropization, we show in Figure~\ref{scattest4} the initial and final energy spectra for two different angles in the fluid-rest frame. As demonstrated evidently in this figure, the initially different spectra converge at the end, implying that neutrino distributions become isotropic at all energies. In Figure~\ref{scattest5} we show the same evolution in the laboratory frame, which is actually the frame we use for simulations. Contrary to the previous case, the initially identical spectrum for different angles is separated as time passes in the laboratory frame, indicating that the final distribution is anisotropic in this frame as it should.

\subsection{Collision term: emission, absorption and isoenergetic scattering combined} \label{subsec:colisoenesca_emiabs}

\begin{figure*}
\vspace{15mm}
\epsscale{0.8}
\plotone{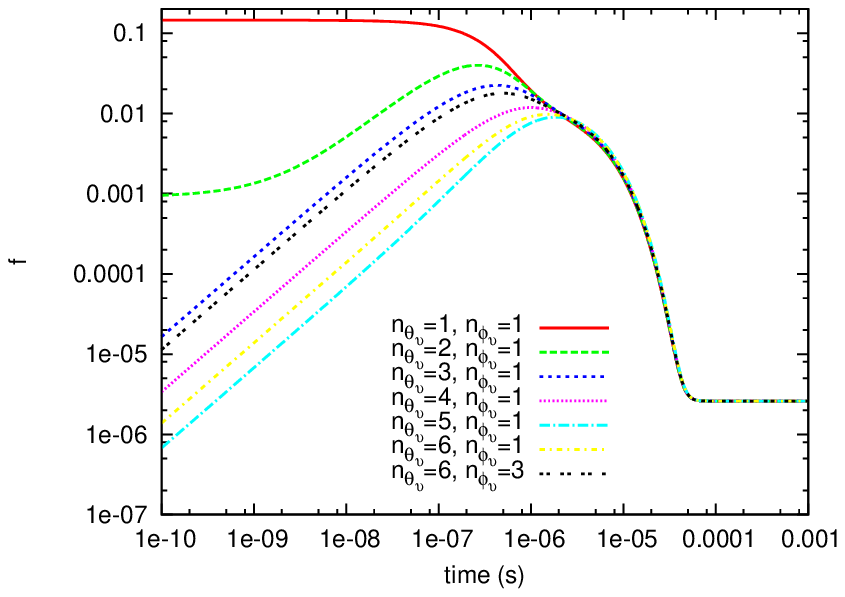}
\caption{The same as Figure~\ref{scattest1} but including emission and absorption processes.
\label{allreactiontest1}} 
\end{figure*}

To the isoenergetic scatterings, we add emissions and absorptions on nucleons and nuclei:
\begin{eqnarray}
e^{-} + p &\longleftrightarrow& \nu_{e} + n , \\
e^{+} + n &\longleftrightarrow& \bar{\nu_{e}} + p , \\
e^{-} + A &\longleftrightarrow& \nu_{e} + A^{'} .
\end{eqnarray}
The initial condition and computational set-up are the same as those in the previous test.

Figure~\ref{allreactiontest1} shows the evolution of $f$ for different angles but with the same energy ($\varepsilon^{\rm{fr}}=60$~MeV) in the fluid rest frame, which corresponds to Figure~\ref{scattest1}. At first the isoenergetic scatterings isotropize the distribution in the fluid-rest frame just as in the previous case. By the time $t \sim 10^{-5}$s, the neutrino distribution is almost isotropic. Note that it is not a Fermi-Dirac distribution at this time, i.e., neutrinos have not yet achieved chemical equilibrium with matter via emissions and absorptions. It is eventually established at $t \sim 10^{-4}$s for this energy of neutrinos. Neutrinos with other energies undergo similar evolutions and reach Fermi-Dirac distributions at different times. It is stressed again that this computation is done on {\it LRG} and the distribution in the fluid-rest frame is obtained via a Lorentz transformation.

\subsection{Advection term: 1D advection through a discontinuity in matter} \label{subsec:trasfer1D}

\begin{figure*}
\vspace{15mm}
\epsscale{0.5}
\plotone{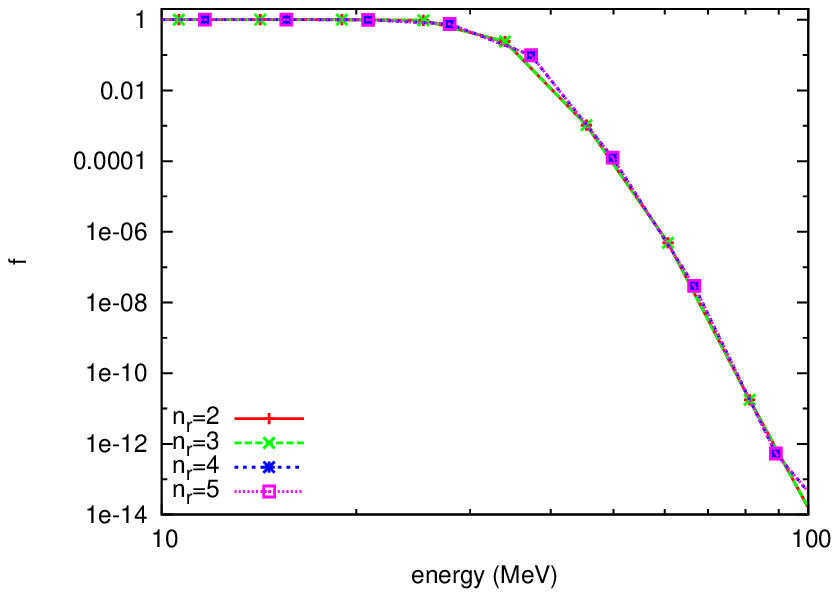}
\plotone{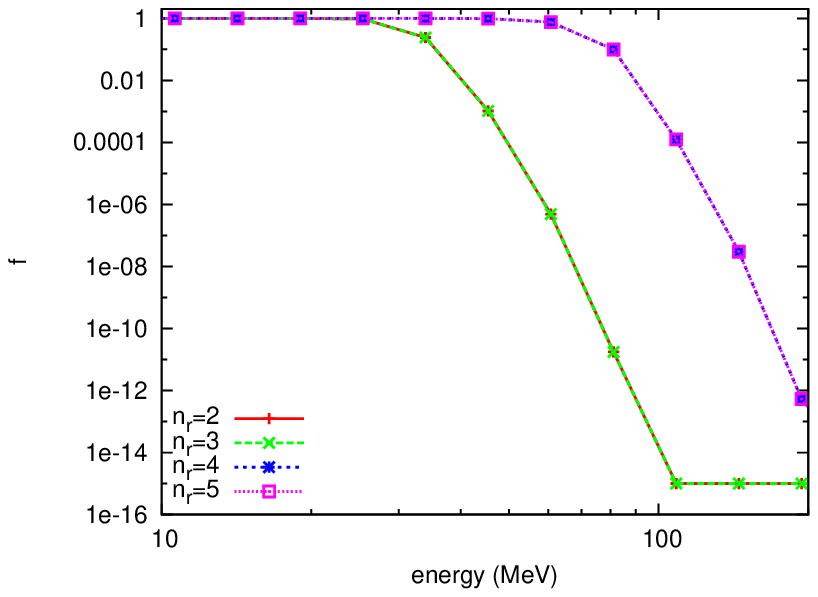}
\caption{Energy spectra of out-going neutrinos at different radii in the vicinity of the discontinuity in the laboratory frame (left) and in the fluid-rest frame (right). $n_{r}$ specifies the radial grid point. Note that the floor value of $f$ is set to be $10^{-15}$.
\label{trasfertest1dvelodisco}} 
\end{figure*}

We now turn to the advection term. Note that this is the main source of difficulties in the ordinary Lagrangian methods. Our formulation treats the spatial and angular advections in the laboratory frame but employs interpolations between two grids ({\it LRG} and {\it LFG}) as detailed in Section~\ref{subsec:step4}. It is hence important to confirm that the scheme indeed works properly.

 Here, we consider the advection in matter that has a discontinuous velocity distribution. This is certainly the most difficult situation for our method. In contrast to the previous tests,  we cut off all neutrino-matter reactions, assuming that the matter is optically thin, and consider the advection term alone. Note that in this case, the energy spectrum of neutrinos is unchanged as they propagate through the discontinuity in the laboratory frame whereas it undergoes a discontinuous change there in the fluid-rest frame (see also Figure~\ref{Lagremap_com_Labo}). We further assume spherical symmetry in this test, i.e., we omit derivative terms with respect to $\theta$, $\phi$, and $\phi_{\nu}$ in Eq.~(\ref{eqn:eqtransfin-spherical}). 


The computational set-up is as follows. To avoid geometrical complications, we consider the advection in a wafer-thin spherical shell: the computational domain covers the range of $ 10^{8} < r < (10^{8} + 10^{5}) $~cm by a uniform radial grid of 6 bins. The matter velocity is piecewise constant with a discontinuity between the 3rd and 4th grid points: v=0 for the first 3 grid points and $v=-2 \times 10^{10} {\rm cm}/ {\rm s}$ for the rest of grid points. These velocities are again fixed during the computation. We inject out-going neutrinos from the radial inner boundary with the Fermi-Dirac distribution that is the same as in the previous tests and follow the subsequent evolution until a steady state is obtained. We deploy an {\it LRG} of $N_{\epsilon}=20$ and $N_{\theta_{\nu}}=6$.

Figure~\ref{trasfertest1dvelodisco} shows that the energy spectra for out-going neutrinos ($n_{\theta_{\nu}}=6$) in the vicinity of the velocity discontinuity in the laboratory frame (left panel) and in the fluid-rest frame (right panel). As is expected, neutrinos advect without any change of their spectrum when they pass through the discontinuity in the laboratory frame. We can also see in this figure that the energy bins for the outer two radial grid points are shifted from those for the inner two. The right panel shows the same spectrum but observed in the fluid-rest frame. Due to the negative radial velocity in the outer region, neutrinos are blue shifted there (see also Figure~\ref{Lagremap_com_Labo}). As demonstrated clearly in this test, our new formulation can reproduce the results just as expected without any numerical problems.


\subsection{Advection term: 3D advection} \label{subsec:trasfer3D}

\begin{figure*}
\vspace{15mm}
\epsscale{0.5}
\plotone{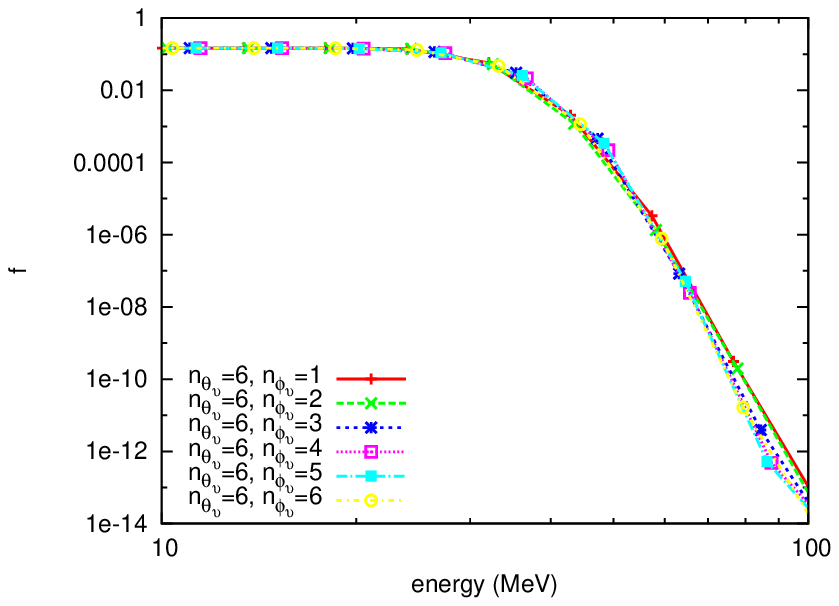}
\plotone{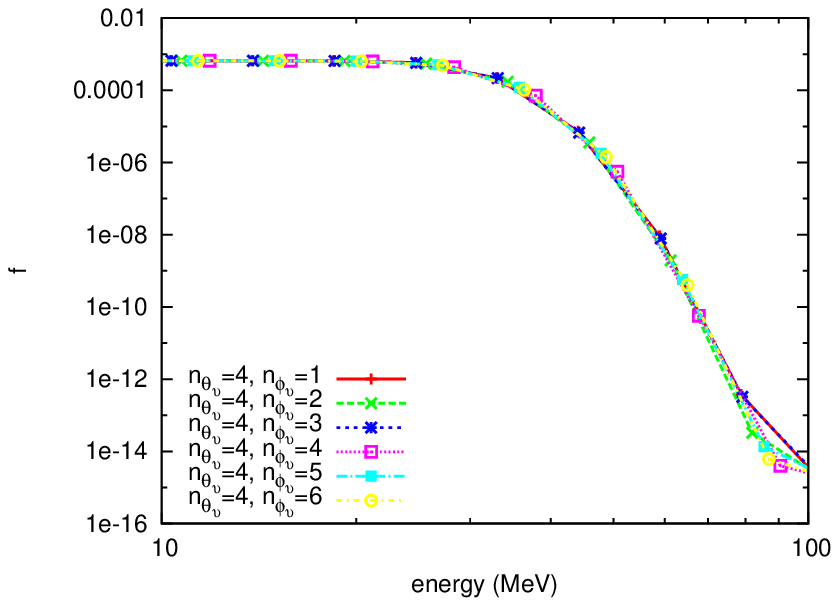}
\plotone{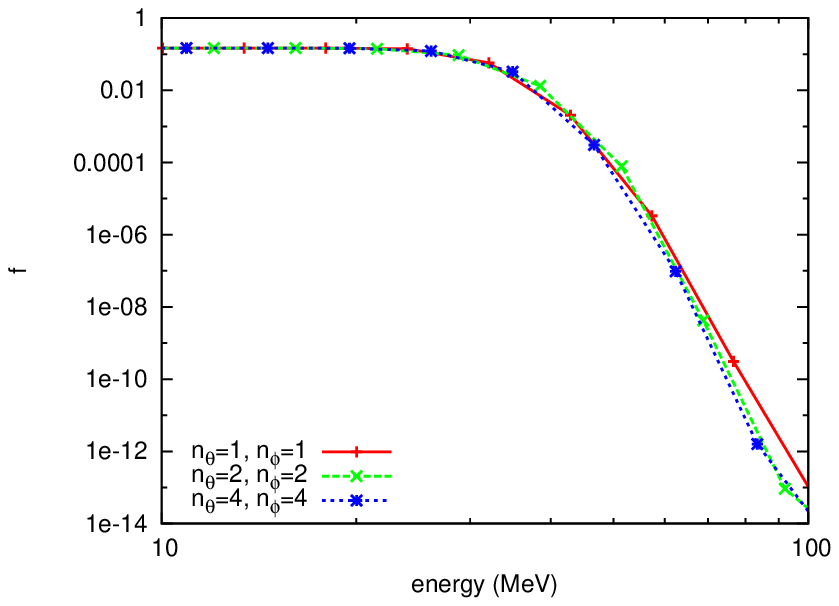}
\caption{Energy spectra at different points in phase space. Upper left: $n_{r} = 6, n_{\theta}=1$, $n_{\theta_{\nu}}=6$ and different $n_{\phi_{\nu}}$. Upper right: the same as the upper left but for $n_{\theta_{\nu}}=4$. Bottom panel: the same as the previous two panels but for $n_{r} = 6, n_{\theta_{\nu}}=6$, and $n_{\phi_{\nu}} = 1$.
\label{3dtransfertest}} 
\end{figure*}

This test is meant to check the Multi-D advection in the optically thin matter with an inhomogeneous non-radial velocity distribution. We assume that the neutrino distribution is spherically symmetric in space. This is no problem, since the matter is optically thin and there is no interaction between the matter and neutrinos. This poses a challenge in our method, however, since {\it LRG} is not spherically symmetric in space and, as a consequence, there is no guarantee that the neutrino distribution remains spherically symmetric in our formulation. This test is hence good diagnostics on our handling of the spatial advection.

The 3D velocity distribution is set in a similar way to that in the previous test, Eq.~(\ref{eq:vdistri_forcol}), but with an additional spatial dependence:
\begin{eqnarray}
v^{r} (r,\theta,\phi) &=& v(r,\theta,\phi) \hspace{1mm} {\rm cos}~\theta_{h}, \\
v^{\theta} (r,\theta,\phi) &=& v(r,\theta,\phi) \hspace{1mm} {\rm sin}~\theta_{h} \hspace{1mm} {\rm cos}~\phi_{h}, \\
v^{\phi} (r,\theta,\phi) &=& v(r,\theta,\phi) \hspace{1mm} {\rm sin}~\theta_{h} \hspace{1mm} {\rm sin}~\phi_{h}. \label{eq:vdistri_for3Dtrans}
\end{eqnarray}
We again set a non-vanishing non-radial velocity by choosing $\theta_{h}=\pi/4$, and $\phi_{h}=\pi/4$. $v(r,\theta,\phi)$ is given as follows:
\begin{eqnarray}
v(r,\theta,\phi) &=& 2 \times 10^{10}  {\rm cos}~A_{r}(r) \nonumber \\
&&  \times {\rm cos}~\theta  \hspace{1mm}  {\rm cos}~\phi \hspace{6mm} ({\rm cm}/{\rm s}), \nonumber \\
A_{r}(r) &=& 2 \pi \times \frac{r - r_{\rm{min}}}{r_{\rm{max}} - r_{\rm{min}}},
   \label{eq:vfix_for3Dtrans}
\end{eqnarray}
where $r_{\rm{max}}$ and $r_{\rm{min}}$ denote, respectively, the maximum and minimum radii of the computational region, which is the spherical shell with $r_{\rm{min}} = 10^{8} < r < r_{\rm{max}} = 10^{8} + 10^{9}$cm, $0 < \theta < \pi$ and $0 < \phi < 2 \pi$. We deploy to this computational domain an {\it LRG} with $N_{r}=6, N_{\theta}=4, N_{\phi}=6, N_{\epsilon}=20, N_{\theta_{\nu}}=6, N_{\phi_{\nu}}=6$. In the following we demonstrate that the neutrino distribution remains spherically symmetric with this small number of spatial and angular grid points. We inject from the inner boundary out-going neutrinos with the Fermi-Dirac distribution employed in the previous tests. The simulation is continued until the neutrino distribution becomes steady.

We summarily display the results of this test in Figure~\ref{3dtransfertest}. The upper left panel shows the energy spectra for different $n_{\phi_{\nu}}$'s (with $n_{\theta_{\nu}}=6$ being fixed) at $n_{r}=6$ and $n_{\theta}=1$ in the laboratory frame. Note that if the neutrino distribution is exactly spherically symmetric, these spectra should coincide with each other. As seen in this figure, they agree quite well despite they are computed on the {\it LRG}, which is not spherically symmetric. The upper right panel is the same as the upper left, but for $n_{\theta_{\nu}}=4$. Note that these neutrinos propagate in a non-radial direction. Again their spectra depend on $\phi_{\nu}$ very weakly. Finally, we display in the bottom panel the energy spectra at a different radial location $n_r=6$. This time, $n_{\theta_{\nu}}=6$ and $n_{\phi_{\nu}} = 1$ are fixed and $n_{\theta}$ is varied. We can confirm also in this case that all energy spectra are in good agreement. It is emphasized again that these results are not trivial and, in fact, the test is very severe, since we assume here very fast matter motions ($\sim 60 \%$ of the speed of light) with large inhomogeneities. We hence think that our new method works satisfactorily.

\subsection{SR Boltzmann-Hydro Simulations: the spherical collapse of $15 M_{\sun}$ progenitor} \label{subsec:1DSRBoltz_Hydro}

\begin{figure*}
\vspace{15mm}
\epsscale{0.9}
\plotone{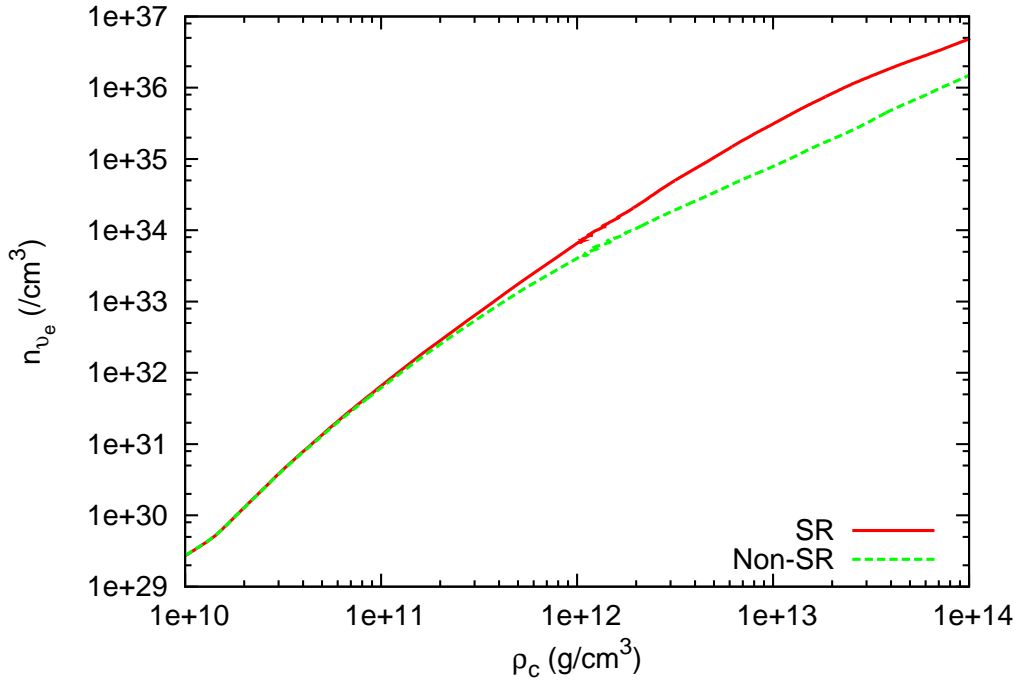}
\caption{The evolutions of the number density of electron-type neutrino at the center both in the SR (red solid line) and NR (green dashed line) simulations. The central matter density instead of the time is used to parametrize the evolutions.
\label{neutrinonumberevo}} 
\end{figure*}

\begin{figure*}
\vspace{15mm}
\epsscale{0.5}
\plotone{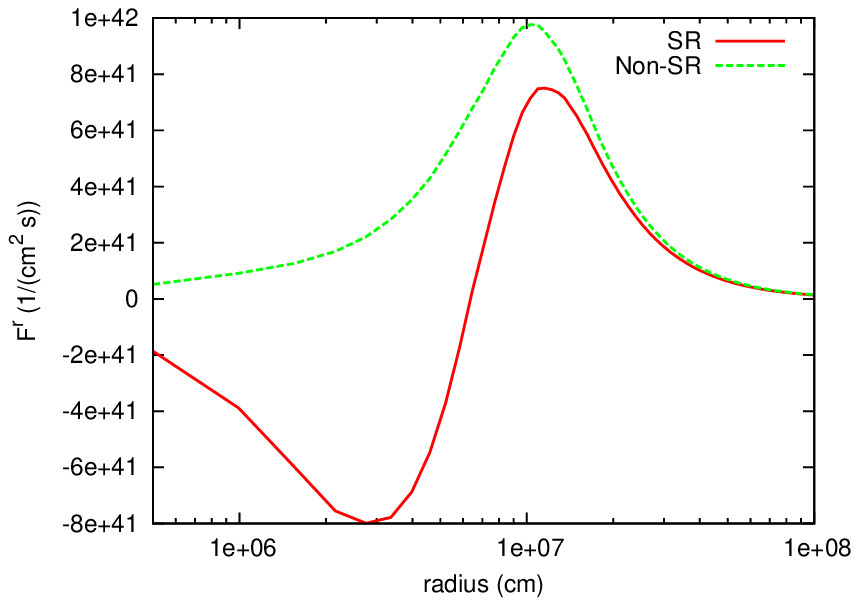}
\plotone{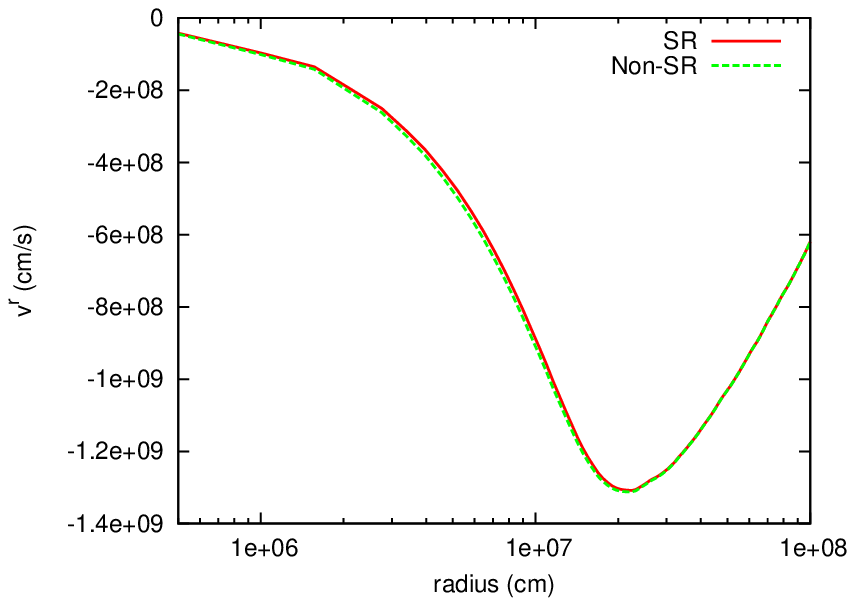}
\plotone{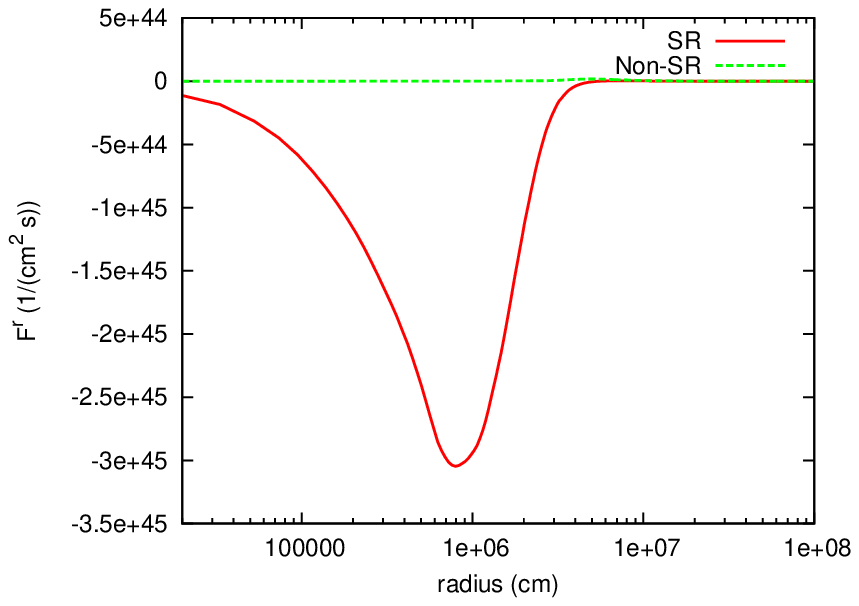}
\plotone{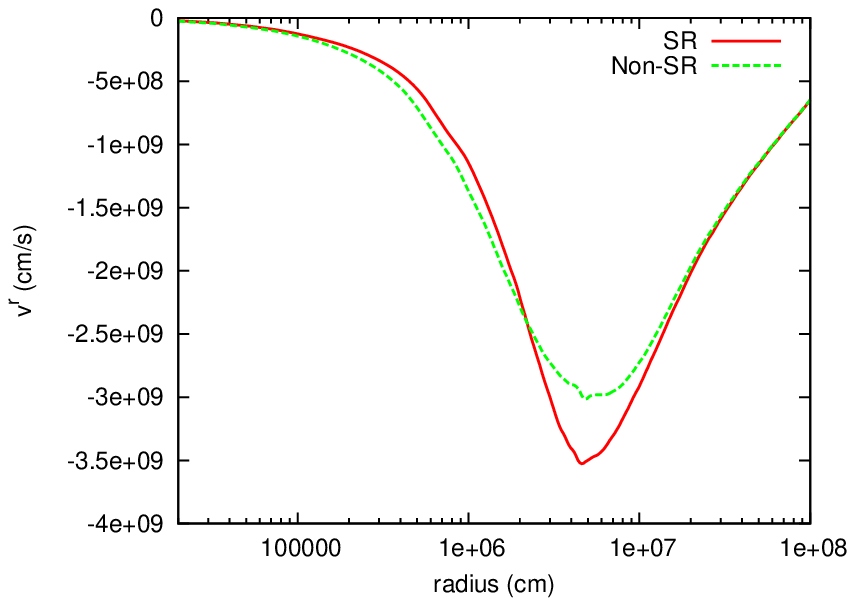}
\caption{The radial distributions of the radial $\nu_{e}$ number flux in the laboratory frame (left panels) and those of the radial matter velocity (right panels). Upper panels show the results at the time when the central density $\rho_{c}$ reaches $10^{12} {\rm{g/cm^3}}$ whereas lower panels correspond to the time of $\rho_{c}=10^{14} {\rm{g/cm^3}}$. The red (green) lines are obtained in the SR (NR) simulations.
\label{neutrinofluxcompare}} 
\end{figure*}


\begin{figure*}
\vspace{15mm}
\epsscale{0.5}
\plotone{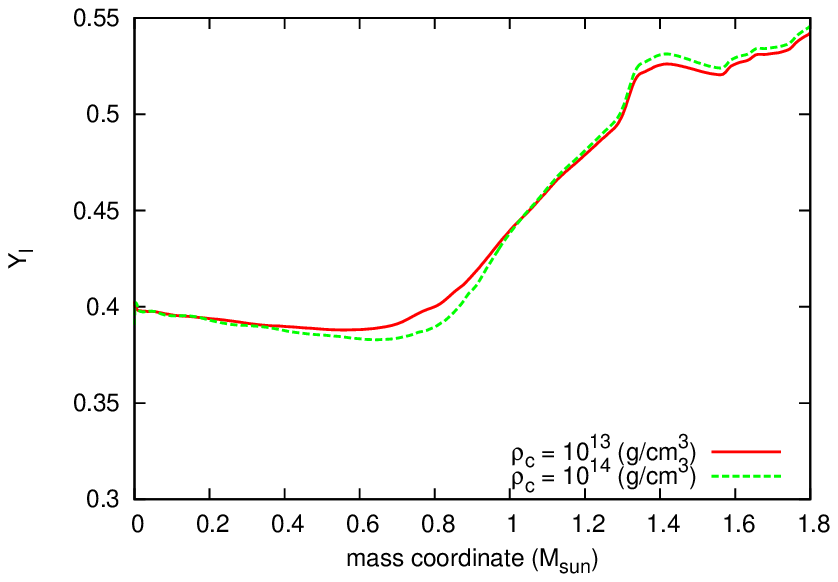}
\plotone{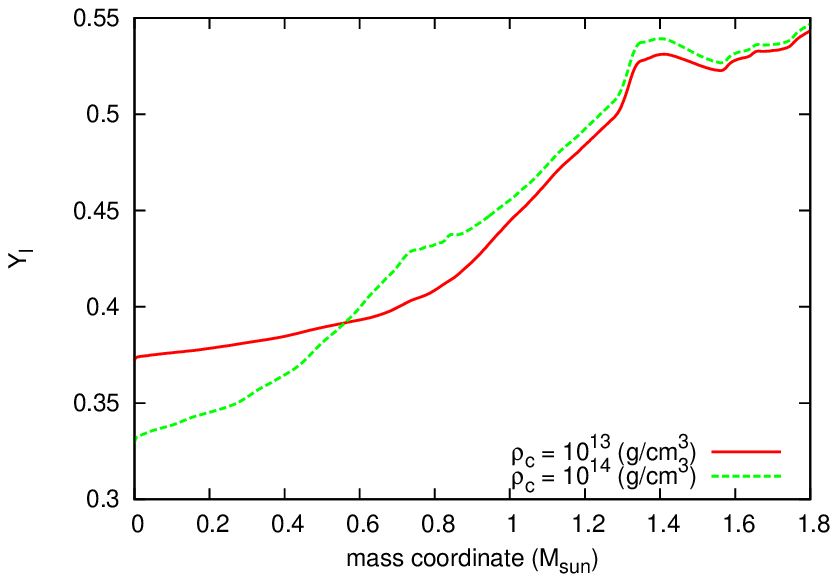}
\caption{The distributions of $Y_{l}$ at two different times. The red (green) lines correspond to the time when the central density reaches $\rho_{c} = 10^{13} (10^{14}) {\rm{g/cm^3}}$. The left panel shows the result in the SR simulation and the right panel is the NR counterpart.
\label{Ylevolutioncompare}} 
\end{figure*}

\begin{figure*}
\vspace{15mm}
\epsscale{1.0}
\plotone{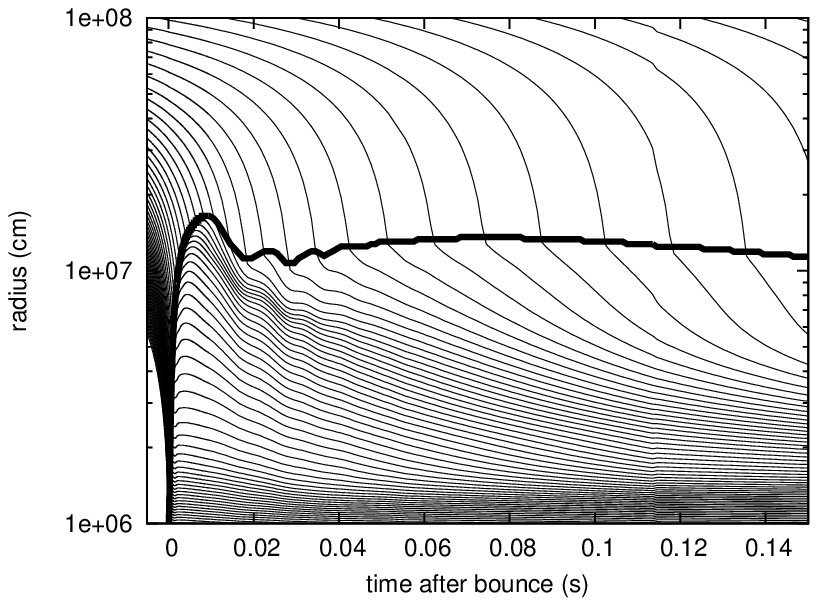}
\caption{Mass shell trajectories in the post-bounce phase. The thick line shows the trajectory of shock wave.
\label{shmassshell}} 
\end{figure*}

So far we have tested the advection and collision separately in simplified set-ups. In reality, however, they are non-linearly coupled with each other and dictate the neutrino transfer and, as a consequence, the dynamics of CCSNe. In order to confirm that our new method is indeed applicable to realistic simulations of CCSNe, we conduct here a 1D spherically symmetric Boltzmann-Hydro simulation for the collapse of $15 M_{\sun}$ progenitor (a non-rotating star with the solar metallicity referred to as s15.0 in \citet{2002RvMP...74.1015W}). We employ an {\it LRG} with $N_{r}=300, N_{\epsilon}=20, N_{\theta_{\nu}}=8$ covering the computational domain of $0<r<4~\times~10^8$cm. For comparison, we also perform a NR simulation for the same set-up. Although the simulation is continued after bounce until the shock wave is stalled, we focus here on the collapsing phase, since the infall velocity is largest and SR effects are most clearly discernible.

Figure~\ref{neutrinonumberevo} shows that the evolution of the number density of $\nu_{e}$ at the center for both the SR and NR simulations. Initially these two simulations follow almost the same evolutionary path. After the central density reaches $\rho_{c} \sim 10^{12} {\rm{g/cm^3}}$, however, they start to deviate and become different by more than a factor of $\sim 4$ at $\rho_{c} \sim 10^{14} {\rm{g/cm^3}}$. During the latter period, neutrinos undergo isoenergetic scatterings on nuclei called coherent scatterings and, as shown shortly, this is the source of discrepancy in fact.

In order to clearly see the SR effects by the matter motion, the left panels of Figure~\ref{neutrinofluxcompare} show as a function of radius the radial component of the number flux, i.e., the energy-integrated first-angular moment of $f_{\nu_{e}}$ in the laboratory frame:
\begin{eqnarray}
F^{r} (r) \equiv \int {{\rm cos}~\theta_{\nu}} f(r,\Omega^{\rm{lb}}, \varepsilon^{{\rm lb}}) d \Omega^{\rm{lb}} d V_{\varepsilon}^{\rm{lb}},  \label{e1:firstmoment}
\end{eqnarray}
where $d V_{\varepsilon}^{\rm{lb}}$ denotes the volume element of energy space in the laboratory frame. The upper panel corresponds to the time when the central density reaches $\rho_{c} = 10^{12} {\rm{g/cm^3}}$ whereas the bottom one shows the result at the time of $\rho_{c} = 10^{14} {\rm{g/cm^3}}$, respectively. On the right panels, matter velocities are displayed as a function of time for the same times.

As is evident in the left panels, the number flux behaves qualitatively differently in the SR and NR cases: $F^{r}$ in the SR simulation is negative in the inner region ($r \lesssim 60$km), whereas it is positive everywhere in the NR. Simply put, neutrinos are moving in the opposite direction if SR is ignored. This is understood as follows (see also Section~\ref{sec:neutrinotrap}): matter is optically thick to neutrinos in the inner region and neutrinos tend to diffuse outwards as observed in the NR simulation; the matter is infalling, on the other hand, and tends to drag neutrinos inwards; this is made possible by frequent interactions between the matter and neutrinos; in fact, as demonstrated in Sections~\ref{subsec:colisoenesca} and \ref{subsec:colisoenesca_emiabs}, scatterings and emissions/absorptions render the neutrino distribution isotropic in the fluid-rest frame and, as a consequence, produce a flux in the direction of velocity in the laboratory frame after Lorentz transformation; if SR is neglected, neutrinos are isotropically distributed even in the laboratory frame and no dragging occurs; this is the cause for the discrepancy. Note that this dragging (and hence SR) is crucially important for neutrino trapping as shown next.

In Figure~\ref{Ylevolutioncompare}, we display the radial distribution of lepton fraction at two different times when the central density reaches $\rho_{c} = 10^{13} {\rm{g/cm^3}}$ and $10^{14} {\rm{g/cm^3}}$. The left panel presents the results of the SR simulation, while the right panel gives the NR counterpart. We can immediately recognize a remarkable difference. In the SR simulation, two lines are almost the same, in particular for $M_{r} < 0.6 M_{\sun}$, where $M_{r}$ denotes the mass coordinate. This means that the lepton number is conserved in each fluid element as it should after neutrino trapping. For the NR case, on the other hand, the lepton fraction is decreased even in the central region, while it is increased in the outer region. This means that neutrinos are diffusing outwards in the Lagrangian frame even after neutrino trapping, which is consistent with what we observed in the number flux above.

In the Lagrangian method, the lepton-number conservation after neutrino trapping is handled almost trivially. Our Boltzmann-Hydro solver is based on the Eulerian picture, in which $Y_{l}$ does evolve as a function of radius even after neutrino trapping. Only when SR is taken into account appropriately, can we reproduce the correct evolutions. The results of this test simulation are hence the clearest evidence that our new method properly handles the neutrino advection. Incidentally, we have also made a comparison with the result of 1D Lagrangian GR simulations \citep{2005ApJ...629..922S}\footnote{In the comparison, we turn off the electron scattering in the Lagrangian GR simulation. Note that GR effect is negligible for $Y_{l}$ before bounce.} and confirmed reasonable agreement between them (although not shown here).


Finally, we present the mass shell trajectories during the post-bounce phase (until 150 ms after the bounce) in Figure~\ref{shmassshell}. After bounce, the shock wave propagates outwards initially through optically thick matter and generates a neutronization burst of $\nu_{e}$ when it breaks out of the neutrino sphere and is eventually stagnated at a certain radius in optically thin matter. These phases are, in general, difficult numerically for our method, since neutrino distributions evolve quite rapidly, and both optically thick and thin regions are involved, and a shock wave, i.e. a discontinuity in matter velocities, exists. In spite of these difficulties, our SR Boltzmann-Hydro code has run stably without problems. Although we have to wait for more detailed quantitative analyses of this model and others in multi-D, which will be reported elsewhere as a sequel, the results shown so far indicate that our new code will be applicable to realistic CCSNe simulations.

\section{Summary and possible extensions of the formulation} \label{sec:summary}
In this paper we have presented a novel method to solve numerically the SR Boltzmann equation in the laboratory frame based on the $S_{n}$ method, which overcomes technical difficulties inherent to the conventional approaches irrespective of the Lagrangian or Eulerian pictures. Our method is hybrid, deploying the Lagrangian remapped grids in the Laboratory frame. The employment of {\it LRG} simply solves the difficulties in the treatment of scatterings, which plague the conventional Eulerian approaches. As a trade-off, the numerical treatment of the advection term becomes complicated as in the ordinary Lagrangian approaches. This problem is mitigated by the use of {\it LFG}, which is nothing but the ordinary grid fixed to the laboratory frame and adopted in the conventional Eulerian approaches. The advection becomes simplest on {\it LFG}. We have developed a scheme for the interpolation between {\it LRG} and {\it LFG}, which ensures the neutrino number conservation.

By carrying out a series of code tests, we have demonstrated that our new method works as expected, correctly handling both the collision and advection terms. With the same code, we have also conducted a 1D CCSNe simulations from core collapse through bounce till shock stall for a realistic progenitor model of $15 M_{\sun}$ with the minimal set of microphysics. We have paid particular attention to the collapsing phase, in which matter velocities reach maximum and our code faces the greatest challenge. We have found that the neutrino-dragging due to matter motions, which is crucially important in neutrino trapping, is correctly captured in the SR simulation but not in the NR one. We have also observed only in the SR computation that the lepton fraction as a function of the Lagrangian mass coordinates does not change in time in the optically thick region. These results clearly indicate that the adequate treatment of SR effects is critically important to obtain the lepton fraction correctly.

 The simulation was continued until the shock wave generated at bounce is stalled in the core. We have found no problem in the later phase, either, and we are now confident that our new method is applicable to the realistic simulation of CCSNe. In fact, we have already started such simulations in 2D and their results will be reported together with further tests in multi-D in our forthcoming paper. It is finally stressed that our method could be applied to other more relativistic phenomena such as photon transfer in AGN or GRBs, since SR effects are taken into account to all orders of $v/c$ in our Boltzmann code. These possibilities will be studied in the future.

At the very end of the paper we comment on the extension of our formulation to GR Boltzmann-Hydro simulations. We have recently published a paper on the conservative form of GR Boltzmann equation \citep{2014PhRvD..89h4073S}, which, in flat space time, would reduce to the one used in the current study. It turns out that our Lagrangian remapping method can be extended to this form of GR Boltzmann equation with some modifications. As shown in Eq.~(21) of that paper, GR modifies only the advection terms with the collision terms being essentially unchanged from the SR case. In the GR case, the choice of {\it LFG} is non-trivial. We may be able to use a local tetrad with a time-like unit vector, $n^{a}$, orthogonal to the spatial hypersurface of $t=$const. Then one important difference from the SR case is that $n^{a}$ depends on space and time, which implies that the GR Boltzmann equation has energy-derivative terms on the left hand side even in the laboratory frame, which is nothing but gravitational redshift. It is noted, however, these terms may not pose problems, since gravitational field change only gradually both in time and space. Such extension is currently underway and will be published elsewhere.







\acknowledgments 
We are grateful to Y. Sekiguchi, A. Mezzacappa, W. Iwakami and S. Furusawa for valuable comments on the multi-D neutrino transfer simulations. We also thank A. Imakura, T. Sakurai and H. Matsufuru for the development of the matrix solver for the Boltzmann equation. The numerical computations were partially performed on the supercomputers at high Energy Accelerator Research Organization (KEK) under the support of its Large Scale Simulation Program (13/14-10). We acknowledge the supercomputing resources at Yukawa Institute for Theoretical Physics (YITP) in Kyoto University. This work was supported by Grant-in-Aid for the Scientific Research from the Ministry of Education, Culture, Sports, Science and Technology (MEXT), Japan (22540296, 24103006, 24740165, 24244036) and HPCI Strategic Program of Japanese MEXT and K computer at the RIKEN (Project ID: hpci 130025, 140211).

\end{document}